# NPV, IRR, PI, PP, and DPP: a unified view

Mikhail V. Sokolov[a,b,c,d]

[a] European University at St. Petersburg, 6/1A Gagarinskaya st., St. Petersburg, 191187, Russia
[b] Centre for Econometrics and Business Analytics (CEBA). St. Petersburg State University, 7/9 Universitetskaya nab., St. Petersburg, 199034, Russia
[c] Institute for Regional Economic Studies RAS, 38 Serpukhovskaya st., St. Petersburg, 190013, Russia
[d] HSE University, 16 Soyuza Pechatnikov st., St. Petersburg, 190121, Russia

**Abstract** This paper introduces a class of investment project's profitability metrics that includes the net present value (NPV) criterion (which labels a project as weakly profitable if its NPV is nonnegative), internal rate of return (IRR), profitability index (PI), payback period (PP), and discounted payback period (DPP) as special cases. We develop an axiomatic characterization of this class, as well as of the mentioned conventional metrics within the class. The proposed approach offers several key contributions. First, it provides a unified interpretation of profitability metrics as indicators of a project's financial stability across various economic scenarios. Second, it reveals that, except for the NPV criterion, a profitability metric is inherently undefined for some projects. In particular, this implies that any extension of IRR to the space of all projects does not meet a set of reasonable conditions. A similar conclusion is valid for the other mentioned conventional metrics. For each of these metrics, we offer a characterization of the pairs of comparable projects and identify the largest set of projects to which the metric can be unequivocally extended. Third, our study identifies conditions under which the application of one metric is superior to others, helping to guide decision-makers in selecting the most appropriate metric for specific investment contexts.

**JEL classification** G11, G31

## 1. Introduction

The most common capital budgeting techniques include the net present value (NPV) criterion, internal rate of return (IRR), payback period (PP), discounted payback period (DPP), and profitability index (PI). Although the capital budgeting literature seems to agree that the NPV criterion outperforms others as an investment criterion, being convenient numerical representations of various aspects of an investment project, the other metrics continue to enjoy widespread use in practice (Siziba and Hall, 2021; Graham, 2022). Moreover, in some cases calculation of a particular metric is required by law.[1] This paper aims to provide a unified perspective on these seemingly different performance measures as profitability metrics.

---

E-mail address: mvsokolov@eu.spb.ru
I am grateful to Alexander Alekseev, Maxim Bouev, Liudmila Galliulina, Alexander Nesterov, Ekaterina Polyakova, and Evgeny Zalyubovsky for helpful comments that greatly improved the paper.
This work was supported by HSE University (Basic Research Program).

[1] Many countries impose restrictions on lending rates in the form of an interest rate cap. In most cases, the effective (rather than nominal) interest rate is restricted, which includes all commissions, fees, and other

The paper employs an axiomatic approach to characterize project's profitability metrics. The characterized metrics include, as a proper subset, a particularly appealing class possessing a multi-utility representation; each member of this representation corresponds to the NPV criterion with a particular discount rate (given a discount rate, the NPV criterion labels an investment project as weakly profitable if its NPV is nonnegative). Elements of this class enjoy a unified interpretation as indicators of a project's financial stability across a set of economic scenarios differing in discount rates. We show that IRR, PP, DPP, and PI belong to this class (Propositions 3, 6, and 8). The multi-utility representation is a straightforward generalization of the one studied in Bronshtein and Akhmetova (2004) and helps shed light on several unanswered questions in the literature. For instance, the literature contains numerous efforts to modify the concept of IRR in order to be well defined for every project. To mention just a few, Arrow and Levhari (1969), Cantor and Lippman (1983), and Promislow and Spring (1996) suggest unconditional solutions, whereas the balance function approach (Teichroew et al., 1965; Spring, 2012), the modified IRR (Lin, 1976; Beaves, 1988; Shull, 1992), the proposal of Hazen (2003), the relevant IRR (Hartman and Schafrick, 2004), the average IRR (Magni, 2010, 2016), and the selective IRR (Weber, 2014) provide solutions conditional on exogenously given reinvestment rate, cost of capital, income, or capital stream. We show that except for a somewhat degenerate case that corresponds to the NPV criterion, a profitability metric is inherently undefined for some projects (Proposition 2). This implies that any extension of IRR, as well as PP, DPP, and PI, to the space of all projects necessarily does not satisfy a set of natural axioms. This result sheds light on the long-standing question of why the conventional metrics – IRR, PI, PP, and DPP – are partial (i.e., are undefined for some projects) and cannot be meaningfully extended to the set of all projects. In the case of IRR, a similar impossibility result was established in Promislow (1997). To overcome this problem, the literature suggests to reduce the space of projects to those for which the profitability metric is well defined (in the case of IRR, e.g., to the space of conventional projects that have only one change of sign in their net cash flow streams). In this paper, we follow a different approach by allowing incomparable projects. This approach helps to determine the largest set of projects to which IRR can be unequivocally extended and, more importantly, offers an incomplete ordinal extension of IRR (Proposition 3). Similar results are provided for the other conventional metrics (Propositions 6 and 8). As a practical implication, we suggest to replace the conventional metrics with their extensions. These extensions enjoy the same natural properties as the underlying metrics, but substantially enlarge the set of comparable projects.

In the capital budgeting literature, it is unanimously accepted that the NPV criterion is the gold standard under certainty and under risk,[2] whereas other profitability metrics such as the rate of return, sometimes put in a subordinated position, may complement the criterion with important pieces of information (e.g., see Magni, 2020, Chapter 7; Hazen and Magni, 2021). Our analysis shows that each metric plays its own role in the presence of (Knightian) uncertainty in project parameters. It suggests that the choice of a particular metric should be determined by the source of

---

expenses associated with a loan (Maimbo and Gallegos, 2014). The regulator, therefore, has to evaluate IRR of the cash flow stream associated with the loan to verify its lawfulness.

[2] In practice, the NPV criterion is also widely used under uncertainty. In real-life applications, decision makers often cannot assess various domain-specific and project-specific factors, in which case a subjectively determined hurdle rate is applied, usually well above the 'correct' cost of capital (Jagannathan et al., 2016; Graham, 2022; Gormsen and Huber, 2023). Such a heuristic has, however, a rigorous justification in terms of bounded rationality (Magni, 2009a) and in the presence of operational constraints (Jagannathan et al., 2016) or beliefs about value creation combined with market power (Gormsen and Huber, 2023).

uncertainty an investor faces, and hence provides a clear condition under which the application of one metric is superior to others. Namely, the NPV criterion should be used under complete certainty. IRR is preferable if the investor faces uncertain discount rate. PP and DPP are superior under uncertain project duration (e.g., if there is a possibility of early project termination for external reasons). PI should be chosen if there is a possibility of reduction (in the form of an unknown scale factor) of future cash flow. While some of these recommendations are already used in practice, our analysis provides a formal justification for them. This approach also provides a clear interpretation for combinations of the metrics. Since there are other sources of uncertainty, the collection of conventional metrics – IRR, PP, DPP, and PI – cannot be considered as comprehensive. As a result, we suggest to replace the zoo of conventional capital budgeting techniques with context-specific profitability metrics determined by sources of uncertainty the investor faces. Construction of such a context-specific profitability metric is illustrated with examples (Section 4.4).

We consider several proper subsets of interest of the space of investment projects (in particular, the set of conventional investments – projects in which a series of cash outflows is followed by a series of cash inflows – and the set of projects that require initial cash outflows) and characterize the structure of profitability metrics defined on them (Lemmas 3–5). This result provides the whole palette of profitability metrics on those sets available to the investor.

Finally, we present characterizations of the conventional profitability metrics. In particular, we show that a profitability metric is well defined for all investment projects requiring an initial outlay if and only if it is consistent with PI (Lemma 5). IRR is known as an extension of the rate of return (the yield rate) defined over the set of investment operations with two transactions (an initial outlay and a final inflow). Although there are other extensions, e.g., the modified IRR and the performance measures introduced in Arrow and Levhari (1969) and Bronshtein and Skotnikov (2007), we show that IRR is a unique one satisfying a set of reasonable axioms (Proposition 3). We also present a genuine counterpart of IRR under nonexponential discounting and provide its axiomatic characterization (Proposition 4). Finally, a profitability metric is well defined for investment operations with two transactions and is stable under truncation (that is, an early termination does not cause a perturbation of the ordering of projects induced by the metric) if and only if it is the DPP obtained using linear interpolation of the cumulative discounted cash flow (Proposition 7). It is a common practice to use linear interpolation to evaluate DPP (e.g., see Götze et al., 2015, p. 72), and our result provides a rigorous justification for it. This justification has nothing to do with the assumption that the observed cash flow is the discretization of a continuous cash flow.

Important contributions to the literature on the axiomatic approach to profitability metrics include Promislow and Spring (1996), Promislow (1997), and Vilensky and Smolyak (1999). In particular, Promislow and Spring (1996) proposed a general measure-theoretic construct for IRR-like profitability metrics. Promislow (1997) is, to our knowledge, the first formal impossibility result that shows that IRR cannot be unambiguously extended to the space of all projects. Vilensky and Smolyak (1999) presented a characterization of IRR as well as of its extensions to nonexponential discounting and stochastic cash flows. An axiomatic approach to valuation of cash flow streams (Norberg, 1990; Promislow, 1994; Spring, 2012) and, more generally, utility streams (Chichilnisky, 1996; Neyman, 2023, to mention just a few) provides significant related results.

The paper is organized as follows. Section 2 attempts to formalize the concept of profitability. It introduces the main object of our analysis, called a profitability ordering, by means of an axiomatic approach. Section 3 studies profitability orderings that are complete being restricted to a given subset of interest. In particular, we characterize profitability orderings that are complete on the set of conventional investments and the set of all investments (projects that require an initial cash outlay). Thus, we describe the general structure of a profitability metric (which we treat as a utility representation of the restriction of a profitability ordering to a subset of interest) defined on those sets. Section 4 shows that various standard capital budgeting metrics, including IRR, PP, DPP, and PI, are profitability metrics, i.e., are induced by profitability orderings. With the help of this result, for each of these metrics we characterize the largest set of projects to which it can be unequivocally extended. All proofs and auxiliary results are given in the Appendix.

## 2. A profitability ordering

We begin with basic definitions and notation. $R_{++}$, $R_+$, and $R$ are the sets of positive, nonnegative, and all real numbers, respectively. $\bar{R} := [-\infty, +\infty]$ and $\bar{R}_+ := [0, +\infty]$ are the extended real and nonnegative real numbers. We equip subsets of $R$ with the usual topology. For a topological space, the topological closure and interior operators are denoted by $cl$ and $int$. The indicator function of a set $S$ is denoted by $I_S$ (that is, $I_S(x) = 1$ if $x \in S$ and $I_S(x) = 0$, otherwise). The one-sided limits and the limit at infinity of a function $f : R_+ \to R$, if they exist and are finite, are denoted by $f(s+) := \lim_{\tau \to s+} f(\tau)$, $s \in R_+$, $f(t-) := \lim_{\tau \to t-} f(\tau)$, $t \in R_{++}$, and $f(+\infty) := \lim_{\tau \to +\infty} f(\tau)$.

Let $P$ be the vector space of real-valued càdlàg functions on $R_+$, that is, $x \in P$ if $x: R_+ \to R$ is right-continuous and has the left limit $x(t-)$ for all $t \in R_{++}$ and the limit $x(+\infty)$. Being endowed with the supremum norm $\|x\| := \sup_{t \in R_+} |x(t)|$, $P$ becomes a Banach space (Monteiro et al., 2018, Theorem 4.2.1, Corollary 4.2.4, pp. 79, 81). Set $P_+ := \{x \in P : x(t) \geq 0 \; \forall t \in R_+\}$ and $P_{++} := \{x \in P : \inf_{t \in R_+} x(t) > 0\}$. Note that $P_+$ is a closed convex cone and $P_{++} = int P_+$. We write $x \geq y$ (resp. $x > y$) if $x - y \in P_+$ (resp. $x - y \in P_{++}$). For any $\tau \in R_+$, denote by $1_\tau$ the function on $R_+$ given by $t \mapsto I_{[\tau, +\infty)}(t)$. Note that $1_0 \in P_{++}$, and therefore, $1_0$ is an order unit for the ordering cone $P_+$ in the vector space $P$ (that is, for every $x \in P$, there is $\lambda \in R_{++}$ such that $\lambda 1_0 \geq x$). An element $x \in P$, called an (investment) *project*, is interpreted as the deterministic cumulative cash flow the project generates, i.e., $x(t)$ is the balance of the project at time $t$ (the difference between cumulative cash inflows and cash outflows over the time interval $[0, t]$).[3] In particular, $P_+$ is the set of projects with the property that the cumulative cash inflow all the time dominates the cumulative cash outflow, and $1_\tau$ is receiving a money unit at time $\tau$. A project is said to be *finite* (resp. *discrete*) if it lies in the linear span of $\{1_\tau, \tau \in R_+\}$ (resp. $\{1_\tau, \tau = 0, 1, ...\}$). It is known (Monteiro et al., 2018, p. 82) that $P$ is the closure of the space of finite projects in the space of bounded functions on $R_+$ endowed with the topology induced by the supremum norm. Thus, $P$ is a natural extension of the practically relevant space of investment projects with finitely many transactions. The topological dual of $P$ is denoted by $P^*$. We equip $P^*$ with the weak$^*$ topology. The dual cone of a set $C \subseteq P$ is defined by $C^\circ := \{F \in P^* : F(x) \geq 0 \ \forall x \in C\}$; the dual cone of a set $K \subseteq P^*$ is defined in a similar fashion, $K^\circ := \{x \in P : F(x) \geq 0 \ \forall F \in K\}$.

---
[3] We prefer to describe a project by means of the cumulative (rather than net) cash flow as this setup enables a uniform treatment of discrete- and continuous-time settings.

Following an axiomatic approach to valuation of cash flow streams (Norberg, 1990; Promislow, 1994), by a *net present value* (*NPV*) *functional*, we mean an additive (i.e., $F(x) + F(y) = F(x+y)$ for all $x, y \in P$) and positive (i.e., $F(P_+) \subseteq R_+$) functional $F : P \to R$ satisfying $F(1_0) = 1$. We let $\mathcal{NPV}$ denote the set all NPV functionals. A routine argument shows that an additive and positive functional on $P$ is homogeneous and continuous (Jameson, 1970, Corollary 3.1.4, p. 81), so $\mathcal{NPV} = \{F \in P_+^\circ : F(1_0) = 1\}$. Denote by $\mathcal{A}$ the set of all nonnegative and nonincreasing functions $\alpha : R_+ \to R_+$ satisfying $\alpha(0) = 1$. It can be shown (see Lemma 8 in the Appendix) that $F \in \mathcal{NPV}$ if and only if there exists $\alpha \in \mathcal{A}$ such that

$$F(x) = x(0) + \int_0^\infty \alpha(t) \mathrm{d}x(t), \qquad (1)$$

where the integral is the Kurzweil-Stieltjes integral.[4] For a finite project $x = \sum_{k=0}^n x_k 1_{t_k}$, where $x_k$ is a net cash flow at time $t_k$, Eq. (1) reduces to the familiar discounted sum $F(x) = \sum_{k=0}^n \alpha(t_k) x_k$. For a continuous project $x(t) = \int_0^t c(\tau) \mathrm{d}\tau$, where $c : R_+ \to R$ is an instantaneous net cash flow, Eq. (1) reduces to another familiar expression, $F(x) = \int_0^\infty \alpha(t) c(t) \mathrm{d}t$. Thus, $F$ generalizes a conventional NPV functional and is able to valuate both discrete and continuous cash flows. As $\alpha(t) = F(1_t)$, i.e., $\alpha(t)$ is the present worth of receiving a money unit (the discount factor) at time $t$, elements of $\mathcal{A}$ are hereafter called *discount functions*.[5] Identities $\alpha(t) = F(1_t)$ and (1) define a one-to-one correspondence between the sets $\mathcal{NPV}$ and $\mathcal{A}$. In what follows, this allows us to identify a discount function with the NPV functional it induces. We use the notation $F^{(\alpha)}$ for an NPV functional whenever we want to emphasize that it is induced by the discount function $\alpha$. The support of a discount function $\alpha$ is denoted by $\mathrm{supp}\{\alpha\} := \{t \in R_+ : \alpha(t) > 0\}$. The discount function that corresponds to the extremely impatient case is denoted by $\chi := I_{\{1\}}$.

Projects profitabilities in this paper are ranked by means of a binary relation $\succeq$ on $P$. The statement $x \succeq y$ means that project $x$ is at least as profitable as $y$. The symmetric and asymmetric parts of $\succeq$ are denoted by $\sim$ and $\succ$. The upper and strict upper contour sets of $\succeq$ at $x \in P$ are $U_\succeq(x) := \{y \in P : y \succeq x\}$ and $U_\succ(x) := \{y \in P : y \succ x\}$. The lower and strict lower contour sets of $\succeq$ at $x \in P$, $L_\succeq(x)$ and $L_\succ(x)$, are defined in a similar fashion. A binary relation $\succeq$ is said to be a *profitability ordering* (*PO* for short) if the following four conditions hold.

*Nontrivial preorder* (*NP*): $\succeq$ is nontrivial (i.e., $\succeq \neq P \times P$), reflexive, and transitive.

*Monotonicity* (*MON*): $x \geq y \implies x \succeq y$.

---
[4] See Monteiro et al. (2018) for a review of the Kurzweil-Stieltjes integral.
[5] Note that in most models involving discounting nonincreasingness of a discount function is an assumption, whereas in our model it is a consequence.

*Internality* (*INT*): for every $x \in P$, the sets $L_\succeq(x)$ and $U_\succeq(x)$ are closed under addition.[6]

*Upper semicontinuity* (*USC*): for every $x \in P$, $U_\succeq(x)$ is closed.

Axiom NP states that projects are comparable in a coherent way. In view of the results on nonextendability of IRR-like profitability metrics to the space of all projects (Promislow, 1997; Vilensky and Smolyak, 1999), we do not require $\succeq$ to be complete (total).[7] Axiom MON ensures that higher cash flow provides higher profitability. Axiom INT states that for any projects $x$, $y$, and $z$, $x \succeq z$ & $y \succeq z$ $\Rightarrow$ $x+y \succeq z$; and $z \succeq x$ & $z \succeq y$ $\Rightarrow$ $z \succeq x+y$. The axiom relates profitability of a pool of projects with profitabilites of its components. In particular, it makes valid the following natural guidance: to guarantee the target level of profitability for a pool of projects, it suffices to achieve the target for each project in the pool. This condition is appealing in practice, since it allows to decompose a complex investment decision into separate evaluations of individual investment projects. Indeed, a manager often has to evaluate the worth of projects that come sequentially. This complicates the decision-making process as the terminal structure of the pool of the undertaken projects is unknown at the stage of making a decision with respect to a particular project. However, assuming that there is a target level of profitability, axiom INT provides a simple guidance: include in the pool only the projects that meet the target. A similar condition plays an important role for the IRR capital budgeting technique: it follows from the definition of IRR that if each project from a pool has higher IRR than the hurdle rate (i.e., is profitable), then so does the IRR, if it exists, of their union. Axiom INT also implies $x \succeq y$ $\Rightarrow$ $x \succeq x+y \succeq y$, that is, the union of a project with a more (resp. less) profitable one increases (resp. decreases) profitability of the union. Again, a similar condition holds for the IRR capital budgeting technique: the IRR of the union of investment projects with the IRRs, say, 5% and 9%, if it exists, is somewhere between 5% and 9%. Axiom INT is closely related to the decomposition axiom used in Promislow (1997) to classify loans by their effective annual rate. A stronger form of the internality axiom was also used in Vilensky and Smolyak (1999, Section 1.4) to characterize IRR. Finally, axiom USC is a regularity condition, which assures that small perturbations of cash flows result in a minor perturbation of the ordering.

The general structure of a PO is described in the following proposition.

**Proposition 1.**

*Conditions NP, MON, INT, and USC are independent (i.e., any three of them do not imply the fourth). For a binary relation $\succeq$ on P the following statements are equivalent:*

(a) $\succeq$ *is a PO;*

(b) *there is a nonempty family $\mathcal{U} \subset 2^{\mathcal{NPV}}$ of nonempty subsets of $\mathcal{NPV}$ such that for all $z \in P$ the set $\bigcap_{K \in \mathcal{U}: z \notin K^\circ} (P \setminus K^\circ)$ is closed under addition and*

$$x \succeq y \iff I_{K^\circ}(x) \geq I_{K^\circ}(y) \text{ for all } K \in \mathcal{U}. \tag{2}$$

*Moreover, without loss of generality, elements of the family $\mathcal{U}$ in part (b) can be chosen to be closed (in the weak* topology) and convex.*

---

[6] A set $C \subseteq P$ is said to be closed under addition if $C + C \subseteq C$.

[7] A binary relation $\succeq$ on P is said to be complete if for all $x, y \in P$, $x \succeq y$ or $y \succeq x$.

Note that the right-hand side of the equivalence (2) can also be represented as $\{K \in \mathcal{U}: x \in K°\} \supseteq \{K \in \mathcal{U}: y \in K°\}$. In what follows, a family $\mathcal{U} \subset 2^{\mathcal{NPV}}$ satisfying the conditions of part (b) of Proposition 1 is called a *representation* of the PO $\succeq$. Clearly, a representation is nonunique. A particular way to choose it is $\mathcal{U} = \{(U_\succeq(z))° \cap \mathcal{NPV}, z \in P\}$.

Further facts about POs are collected in the next lemma.

**Lemma 1.**

*A PO $\succeq$ enjoys the following properties.*

1. $\lambda x \sim x$ *for all* $x \in P$ *and* $\lambda > 0$.
2. $\succeq$ *is not lower semicontinuous (i.e., it is not true that* $L_\succeq(x)$ *is closed for every* $x \in P$ *).*
3. *There are* $x, y \in P$ *such that* $x > y$ *and* $x \sim y$.
4. *There are* $x, y \in P$ *such that* $x \succ y$ *and* $x \sim x + y$ *(similarly, there are* $x, y \in P$ *such that* $x \succ y$ *and* $y \sim x + y$ *).*
5. *If* $1_0 \succ x$ *and* $1_0 \succ -x$, *then* $x$ *and* $-x$ *are incomparable.*
6. *The intersection of a collection of POs is a PO.*

Property 1 states that profitability takes no account of the investment size and hence is a relative measure. All known measures of profitability satisfy this property.

Upper and lower semicontinuity are desirable properties as cash flows contain future components which are measured with an error. However, unless $x \sim 1_0$, we have $0 \notin L_\succeq(x)$, so lower semicontinuity does not hold (property 2).

Even though $x \geq y \Rightarrow x \succeq y$, strictly higher cash flow does not necessarily imply strictly higher profitability (property 3). If $x \succ y$, then one would expect $x \succ x + y \succ y$ (recall that, by INT, $x \succeq x + y \succeq y$). For instance, if two investment projects have the IRRs $r$ and $s$ with $r < s$, then the IRR of their union, if it exists, falls strictly between $r$ and $s$. Unfortunately, strict inequalities cannot hold for every pair of comparable projects (property 4).

It is widely recognized in the literature that ambiguity of the definition of IRR for an arbitrary cash flow is a consequence of a change in the status of the investor from that of a lender to that of a borrower for mixture projects (Teichroew et al., 1965; Gronchi, 1986; Hazen, 2003; Hartman and Schafrick, 2004; Promislow, 2015, Section 2.12; Magni, 2016). Property 5 shows that every profitability metric suffers from the same drawback. The problem is that pure investment and pure financing, no matter how they are defined, differ by sign and, therefore, by property 5, are incomparable.

Finally, property 6 provides a way to aggregate multiple profitability criteria.

A simple but important example of a PO is given by the NPV criterion. A PO $\succeq$ is said to be the *NPV criterion* if there is $F \in \mathcal{NPV}$ such that $x \succeq y \Leftrightarrow I_{\{F\}°}(x) \geq I_{\{F\}°}(y)$. The NPV criterion partitions $P$ into the sets of (nonstrictly) profitable and unprofitable projects with, respectively, nonnegative and negative NPV. Properties of the NPV criterion are collected in the following proposition.

**Proposition 2.**

*For a PO $\succeq$, the following statements are equivalent:*

(a)     $\succeq$ *is the NPV criterion;*

(b)     $\succeq$ *is complete;*

(c)     *for every* $x \in P$, $L_{\succeq}(x)$ *is closed under addition;*

(d)     *for every* $x \in P$, $U_{\succeq}(x)$ *is closed under addition.*

An important consequence of Proposition 2 is that, with the exception of a somewhat degenerate case that corresponds to the NPV criterion, all POs are incomplete. As we will show in Section 4, the conventional capital budgeting metrics, in particular, IRR, PP, DPP, and PI, are utility representations of the restrictions of POs to the domains of the metrics. Proposition 2, therefore, sheds light on the question of why these metrics are partial (i.e., are undefined for some projects) and cannot be meaningfully extended to the set of all projects $P$. For instance, the literature contains numerous efforts to modify the concept of IRR in order to be well defined for every project. Proposition 2 implies that such efforts necessary result in a ranking that does not satisfy the axioms introduced above. The same conclusion holds for the other metrics.

Proposition 2 also shows that unless it is the NPV criterion, a PO has the following undesirable property: the union of projects having strictly lower (resp. higher) profitabilities than a project $x$ does not necessarily produce strictly lower (resp. higher) profitability than $x$.

**Example 1.**

Canada's Criminal Code prohibits lending money at an effective annual rate exceeding 60%, where the effective rate includes all charges and expenses paid or payable for the advancing of credit, with only a few exceptions.[8] Given that the effective rate is the IRR of the cash flow stream associated with the loan, how should this provision be interpreted for loans whose cash flows have no IRR? This problem is studied in detail in Promislow (1997). In this example, we partly follow an axiomatic approach of Promislow (1997) to show how to extend the statement of the provision to all loans.

Let $U \subset P$ be the set of lender's cash flows associated with usurious (illegal) loans. The following conditions on $U$ and $N := P \setminus U$ – the set of lender's cash flows associated with nonusurious (legal) loans – seem to be reasonable.

1. $x \geq y$ & $y \in U \Rightarrow x \in U$.
2. The sets $U$ and $N$ are closed under addition.
3. $N$ is open.
4. $-1_0 + a1_t \in N$ (resp. $-1_0 + a1_t \in U$), whenever $t > 0$ and $a < 1.6^t$ (resp. $a > 1.6^t$).

Condition 1 states that a loan with a higher lender's cash flow than a usurious loan is usurious. An equivalent dual condition asserts that $x \leq y$ & $y \in N \Rightarrow x \in N$. That is, a loan with a lower lender's cash flow than a nonusurious loan is nonusurious. By condition 2, a lender cannot get around the law and make a usurious loan by decomposing it into two nonusurious ones. Similarly, the union of two usurious loans is usurious. According to 3, a minor perturbation of a nonusurious loan is nonusurious. Finally, condition 4 states that simple loans with two transactions – an initial lending and final repayment – having an effective annual rate of lower (resp. higher) than 60% are nonusurious (resp. usurious).

---

[8] Criminal Code, Revised Statutes of Canada 1985, c. C-46, Section 347.

To characterize U, define the binary relation $\succeq$ on P by $x \succeq y \Leftrightarrow I_U(x) \geq I_U(y)$. From conditions 1–3 it follows that $\succeq$ is a complete PO, and, therefore, by Proposition 2, it is the NPV criterion. Condition 4 implies that the NPV functional is induced by the discount function $t \mapsto 1.6^{-t}$. The obtained result suggests to correct the statement of the provision of the Criminal Code and classify a loan with a lender's cash flow $x \in P$ as nonusurious (resp. usurious) if $F_{60\%}(x) < 0$ (resp. $F_{60\%}(x) \geq 0$), where $F_{60\%}$ is the NPV functional associated with the discount function $t \mapsto 1.6^{-t}$. The suggested rule is consistent with the current one: if a lender's cash flow $x$ has the IRR less than (resp. of or exceeding) 60%, then $F_{60\%}(x) < 0$ (resp. $F_{60\%}(x) \geq 0$).

To illustrate the suggested rule consider a 1-year loan of 1 money unit, with an application fee of 2% of the loan amount paid 1 day prior. Assuming that time is measured in years and the loan nominal annual interest rate is $r$, the lender's cash flow is $x = 0.02 \cdot 1_0 - 1 \cdot 1_{1/365} + (1+r) \cdot 1_{366/365}$. Since $x$ starts with an inflow, for any $r > -1$, it does not possess the IRR in the conventional sense (i.e., it is not true that the IRR polynomial has a unique root, and at this root, the polynomial changes sign from positive to negative). Therefore, the current statement of the provision is inapplicable to $x$. This is counterintuitive at least when $r > 60\%$. Indeed, $x$ is more favorable to the lender than the same loan without the application fee, $-1 \cdot 1_{1/365} + (1+r) \cdot 1_{366/365}$. If $r > 60\%$, the latter loan is usurious, and therefore, $x$ is expected to be usurious too. In contrast, the suggested rule is applicable to all cash flows. In particular, it classifies $x$ as usurious if and only if $r \geq 56.8\%$ (as $F_{60\%}(x) \geq 0 \Leftrightarrow r \geq 56.8\%$).

A PO $\succeq$ is said to be *symmetric* (*SPO* for short) if there is a nonempty set $\mathcal{F} \subseteq \mathcal{NPV}$ such that

$$x \succeq y \Leftrightarrow I_{\{F\}^\circ}(x) \geq I_{\{F\}^\circ}(y) \text{ for all } F \in \mathcal{F}. \tag{3}$$

By property 6 in Lemma 1, the binary relation defined by (3) is indeed a PO. Note that the right-hand side of the equivalence (3) can also be represented as $\{x\}^\circ \cap \mathcal{F} \supseteq \{y\}^\circ \cap \mathcal{F}$. The multi-utility representation (3) has a straightforward interpretation. As is well known, evaluation of future cash flows depends crucially on the discount rate or, more generally, the discount function used to evaluate NPV. The discount function is often uncertain, especially for long-term investments (Weitzman, 2001; Chambers and Echenique, 2018). Consider a set of possible economic scenarios differing in discount function used to determine an NPV functional. Let $\mathcal{F}$ be the set of NPV functionals induced by those discount functions. In what follows, we identify a scenario with the element of $\mathcal{F}$ it induces; so $\mathcal{F}$ describes the uncertainty an investor faces (this is the Knightian uncertainty in the sense that the investor knows the set of possible scenarios $\mathcal{F}$, but is unable to quantify their odds).[9] Then, according to (3), $x \succeq y$ if and only if for any scenario $F \in \mathcal{F}$, if project $y$ is profitable, i.e., has nonnegative NPV, $F(y) \geq 0$, then so is project $x$.[10] In other words, $x \succeq y$ if and only if $x$ is profitable in a larger set of scenarios than $y$. The interpretation shows that the SPO measures financial stability of projects across the set of scenarios $\mathcal{F}$. By way of illustration, suppose that the investor faces uncertain cost of capital in the range of 7–11%. Then the SPO $\succeq$ induced by $\mathcal{F} = \{F_\lambda, \lambda \in [0.07, 0.11]\}$, where $F_\lambda$ is the NPV functional associated with the discount function $t \mapsto (1+\lambda)^{-t}$, is an appropriate tool to evaluate project profitability. For instance, for projects $x = -10 \cdot 1_0 + 21 \cdot 1_1 - 11 \cdot 1_2$ and $y = -100 \cdot 1_0 + 109 \cdot 1_1$, we have $x \succ y$ as $\{F \in \mathcal{F} : F(x) \geq 0\} = \{F_\lambda, \lambda \in [0.07, 0.1]\}$ is a proper superset of $\{F \in \mathcal{F} : F(y) \geq 0\} = \{F_\lambda, \lambda \in [0.07, 0.09]\}$, or, in words, project $x$ is profitable when the cost of capital is in the range of 7–10%, whereas project $y$ is profitable only when the cost of capital is in the range of 7–9%.

---

[9] Although uncertainty with respect to discount functions is the only type of uncertainty considered, it can sometimes be equivalently interpreted as uncertainty with respect to cash flows rather than discount functions. For instance, assume that the investor faces the possibility of early project termination in a year for an external reason. This results in uncertainty with respect to cash flows represented by two scenarios – a base one and the scenario that starts as the base one but has zero net cash flow after 1 year. Uncertainty with respect to discount rates is not assumed and we denote by $\lambda > 0$ the discount rate in both scenarios. One can equivalently interpret this situation as uncertainty with respect to discount functions rather than cash flows.

In the case of discrete projects, two particular SPOs (with $\mathcal{F} = \mathcal{NPV}$ and $\mathcal{F}$ being the set of NPV functionals associated with the family of exponential discount functions) were studied in Bronshtein and Akhmetova (2004). In what follows, we mainly exploit SPO due to its simple representation and particularly nice interpretation. In Section 4, we show that IRR, PP, DPP, and PI are utility representations of the restrictions of SPOs to the domains of the metrics, and therefore, the orderings they induce can be represented by means of (3).

We call the set $\mathcal{F}$ in (3) a *representation* of the SPO $\succeq$ (we also say that $\mathcal{F}$ *represents* or *induces* $\succeq$). A representation of an SPO is in general nonunique. For instance, if $F, G \in \mathcal{NPV}$ and W is a dense subset of the interval $(0,1)$, then $\{wF + (1-w)G, w \in W\}$ and $\{wF + (1-w)G, w \in (0,1)\}$ represent the same SPO. In what follows, by *the* representation of an SPO $\succeq$, we mean the largest subset of $\mathcal{NPV}$ representing $\succeq$, that is, $\bigcap\{F \in \mathcal{NPV} : I_{\{F\}^\circ}(x) \geq I_{\{F\}^\circ}(y)\}$, where the intersection is taken over all pairs $x, y \in \mathrm{P}$ such that $x \succeq y$.

The structure of an SPO with a closed and convex representation is described in the following example.

**Example 2.**

Given a nonempty closed (in the weak* topology) set $\mathcal{S} \subseteq \mathcal{NPV}$, let $\mathcal{F}$ be the closed convex hull of $\mathcal{S}$ and $\succeq$ be the SPO induced by $\mathcal{F}$. To motivate the idea behind $\succeq$ note that since $\mathrm{P}_{++} \neq \varnothing$, the set $\mathcal{NPV}$ is compact (Jameson, 1970, Theorem 3.8.6, p. 123) and therefore so is $\mathcal{F}$. As $\mathcal{S}$ is closed, it contains the closure of the set of extreme points of $\mathcal{F}$, and by the integral form of the Krein-Milman theorem (Kadets, 2018, Theorem 2, p. 510), every $F \in \mathcal{F}$ can be treated as the expected NPV under a probability measure over $\mathcal{S}$. Thus, the SPO $\succeq$ induced by $\mathcal{F}$ can be interpreted as a measure of project's financial stability under (unknown) probabilistic uncertainty with respect to the set of scenarios $\mathcal{S}$.

---

Just consider two scenarios with the same cash flow as in the base scenario but with different discount functions, $t \mapsto (1+\lambda)^{-t}$ and $t \mapsto (1+\lambda)^{-t} I_{[0,1]}(t)$. Hereafter, the two interpretations are used interchangeably.

[10] According to the residual income approach, a project is (weakly) profitable if and only if the present value of the residual income is nonnegative. It has been shown in the literature that this approach, based on incomes and capitals in place of cash flows, is equivalent to the conventional NPV criterion (Magni, 2009b). This equivalence provides a dual interpretation of representation (3) based on the present value of the residual income rather than of the cash flow.

One can show (see Lemma 9 in the Appendix) that
$$x \succeq y \iff \sup_{\lambda \in \mathrm{R}_+} \min_{F \in \mathcal{S}} F(x - \lambda y) \geq 0. \tag{4}$$

Moreover, if $\mathcal{S}$ is not necessarily closed, then the minimum in (4) needs to be replaced by the infimum. In particular, if $\mathcal{S}$ is finite, then $x \succeq y$ if and only if there exists $\lambda \in \mathrm{R}_+$ such that $F(x - \lambda y) \geq 0$ for all $F \in \mathcal{S}$. Or, in words, $x \succeq y$ if and only if project $y$ can be rescaled such that $x$ has a nonstrictly higher NPV than the rescaled $y$ in every scenario from $\mathcal{S}$.

By way of illustration, consider the SPO $\succeq$ induced by $\mathcal{NPV}$, i.e., any discount pattern may occur. From representation (4) with $\mathcal{S} = \mathcal{NPV}$ and the identity $\mathcal{NPV}^\circ = \mathrm{P}_+$ (which follows from the bipolar theorem (Aliprantis and Border, 2006, Theorem 5.103, p. 217)) we deduce that $x \succeq y$ if and only if for any $\varepsilon > 0$, there exists $\lambda \in \mathrm{R}_+$ such that $x + \varepsilon 1_0 \geq \lambda y$. In particular, if $x$ and $y$ are finite, then $x \succeq y$ if and only if there exists $\lambda \in \mathrm{R}_+$ such that $x \geq \lambda y$, i.e., project $y$ can be rescaled such that its cumulative cash flow is dominated by that of $x$.

We proceed by describing real-valued functions suitable for profitability measurement purposes. A function $M : \mathrm{Q} \to \mathrm{R}$ defined on a nonempty set $\mathrm{Q} \subseteq \mathrm{P}$ is said to be a *profitability metric* if there exists an SPO $\succeq$ such that for any $x, y \in \mathrm{Q}$, $x \succeq y \iff M(x) \geq M(y)$. An SPO satisfying this property is said to be *M-consistent*. Put differently, a profitability metric is a utility representation of the restriction of an SPO. Being a profitability metric seems to be a minimally reasonable requirement for a performance measure to evaluate profitability. For instance, one can show (see Section 4.1) that the conventional IRR is a profitability metric, whereas the modified IRR (Lin, 1976) is not. Keeping in mind the interpretation of an SPO introduced above, we can also treat a profitability metric $M$ as an indicator of a project's financial stability across a set of economic scenarios representing an $M$-consistent SPO.

As shown in Proposition 2, the structure of profitability metrics on P is rather trivial: $I_{\{F\}^\circ}$, $F \in \mathcal{NPV}$ are, up to an order-preserving transformation, the only profitability metrics on P. However, an investor can often be satisfied with profitability metrics defined on a proper subset of P, which are more diverse. For instance, a generic net cash flow stream for investment in a stock is composed of an initial outlay associated with buying the stock, dividends received during the holding period, and the gain that occurs when the stock is sold. Therefore, a profitability metric defined on the set of projects in which an initial capital outflow is followed by a series of cash inflows is a reasonable tool to evaluate stock performance. As another example, a loan usually generates a conventional cash flow that has only one change of sign in its net cash flow stream. Thus, a profitability metric defined on the set of conventional cash flows is an adequate tool to evaluate loans. The rest of the paper deals with the following two problems.

*Problem A*: given a set $\mathrm{Q} \subseteq \mathrm{P}$, describe all profitability metrics defined on $\mathrm{Q}$.

*Problem B*: given a function $M : \mathrm{Q} \to \mathrm{R}$, $\mathrm{Q} \subseteq \mathrm{P}$, check whether $M$ is a profitability metric, and if it is, describe the set of all $M$-consistent SPOs.

Given a set of projects $\mathrm{Q}$, solution of problem A provides the whole palette of profitability metrics available to an investor considering projects from $\mathrm{Q}$. In Section 3, problem A is solved for a few particular sets of interest (in particular, the set of conventional investments and the set of all investments – projects that require an initial cash outlay). Section 4 deals with problem B; we show that various standard capital budgeting performance measures, including IRR, PP, DPP, and PI, are profitability metrics and describe the corresponding consistent SPOs. These SPOs are considered as genuine extensions of the metrics: enjoying the same natural properties as the underlying profitability metrics, they substantially enlarge the set of comparable projects.

## 3. Completeness on a predetermined subset of projects

Problem A is to describe all profitability metrics defined on a predetermined subset of projects. This section aims to solve this problem for a few particular sets of interest.

We begin by introducing the following definition. Given $Q \subseteq P$, an SPO is said to be $Q$-*complete* if its restriction to $Q$ is complete (i.e., $Q$ is a chain). Clearly, $Q$-completeness is a necessary condition for an SPO to induce a profitability metric on $Q$. At least when $Q$ is second countable (e.g., this holds in the practically relevant case of discrete projects), this is also a sufficient condition due to a utility representation theorem of Rader (1963). In view of this, when solving problem A, we restrict ourselves with describing the set of all $Q$-complete SPOs.

Given $Q \subseteq P$ and $\mathcal{F} \subseteq \mathcal{NPV}$, we say that a preorder $\geqslant$ on $\mathcal{F}$ is *induced by* $Q$ if for any $F, G \in \mathcal{F}$, $F \geqslant G \Leftrightarrow \{F\}^\circ \cap Q \supseteq \{G\}^\circ \cap Q$. The relations $\succ$ and $\sim$ are defined as usual. The relation $\geqslant$ has a straightforward interpretation. The statement $F \geqslant G$ means that scenario $F$ is more favorable than $G$ for an investor considering projects from $Q$. Indeed, $F \geqslant G$ if and only if each project $x \in Q$ which is profitable under $G$ (i.e., $G(x) \geq 0$) is also profitable under $F$.

The next lemma is a useful tool in verifying $Q$-completeness. It shows that an SPO is $Q$-complete if and only if elements of its representation can be completely preordered in a natural way.

**Lemma 2.**

*Let $Q \subseteq P$, $\succeq$ be an SPO with a representation $\mathcal{F}$, and $\geqslant$ be the preorder on $\mathcal{F}$ induced by $Q$. Then $\succeq$ is $Q$-complete if and only if $\geqslant$ is complete.*

To illustrate Lemma 2, let $Q$ be the set $\{x \in P: x(0) < 0$ and $x$ is nondecreasing$\}$ of projects in which an initial cash outflow is followed by a series of cash inflows, $\mathcal{F}$ be the set of NPV functionals associated with the family of exponential discount functions $\{t \mapsto e^{-\lambda t}, \lambda \in R_+\}$, and $\geqslant$ be the preorder on $\mathcal{F}$ induced by $Q$. It is straightforward to verify that if $F$ and $G$ are the NPV functionals induced, respectively, by $t \mapsto e^{-\lambda t}$ and $t \mapsto e^{-\beta t}$, then $F \geqslant G$ if and only if $\lambda \leq \beta$. Thus, for a project from $Q$, the smaller the exponential discount rate, the more favorable scenario. This conclusion seems to be obvious; however, it does not hold if $Q$ is, say, the set $\{x \in P: x(0) < 0\}$ of all investments that require an initial cash outlay. Clearly, $\geqslant$ is complete, and, therefore, by Lemma 2, the SPO induced by $\mathcal{F}$ is $Q$-complete. This means that $\mathcal{F}$ induces a profitability metric on $Q$; below we show that, up to an order-preserving transformation, this profitability metric is the IRR.

In the rest of this section, we describe the structure of $Q$-complete SPOs for $Q$ comprising various types of investments. These results relate completeness over several notable subsets of projects and completeness of the restriction of well-known partial orderings over $\mathcal{A}$. Put $Q_1 \coloneqq \{x \in P: x(0) < 0$ and $x$ is nondecreasing$\}$, $Q_2 \coloneqq Q_1 \cup \{x \in P \setminus P_+: x(0) \leq 0$ and there is $\tau \in \mathrm{R}_{++}$ such that $x$ is nonincreasing (resp. nondecreasing) on $[0,\tau)$ (resp. $[\tau,+\infty))\}$, $\mathrm{Q}_3 := \{x \in \mathrm{P} : x(0) < 0\}$, and $\mathrm{Q}_4 := \mathrm{Q}_3 \cup \{x \in \mathrm{P} : x(0) = 0$ and there is $\tau \in \mathrm{R}_{++}$ such that $x(\tau) < 0$ and $x$ is nonincreasing on $[0,\tau]\}$. The set $\mathrm{Q}_1$ ($\mathrm{Q}_2$) consists of conventional investments in which an initial cash outflow (a series of cash outflows) is followed by a series of cash inflows. The set $\mathrm{Q}_3$ ($\mathrm{Q}_4$) comprises all investments, i.e., the projects that require an initial cash outflow (outflows).

We endow $\mathcal{NPV}$ (or, equivalently, $\mathcal{A}$) with the two partial orderings. Let $F, G \in \mathcal{NPV}$ and $\alpha$ and $\beta$ be the discount functions associated with $F$ and $G$, respectively. We write $F \succcurlyeq_1 G$ if $\alpha \geq \beta$ (pointwise). The relation $\succcurlyeq_1$ describes the strength of discounting. We write $F \succcurlyeq_2 G$ if $\mathrm{supp}\{\alpha\} = \mathrm{supp}\{\beta\}$ and the function $t \mapsto \alpha(t)/\beta(t)$ defined on $\mathrm{supp}\{\beta\}$ is nondecreasing. Provided that the discount functions are positive and differentiable, $F \succcurlyeq_2 G$ holds if and only if the instantaneous discount rate under $\beta$ dominates that under $\alpha$, $-(\ln \beta)' \geq -(\ln \alpha)'$. $\succcurlyeq_2$ is known as the patience ordering (e.g., see Quah and Strulovici, 2013, Section II.C). Note that $\succcurlyeq_2 \subset \succcurlyeq_1$.

The next two lemmas describe the structure of $\mathrm{Q}_1$- and $\mathrm{Q}_2$-complete SPOs.

**Lemma 3.**

*Let $\succeq$ be an SPO with a representation $\mathcal{F}$. Put $\mathrm{Q}'_1 := \{-1_0 + a1_\tau, a \geq 1, \tau > 0\}$. The following conditions are equivalent:*

(a) $\succeq$ is $\mathrm{Q}_1$-complete;

(b) $\succeq$ is $\mathrm{Q}'_1$-complete;

(c) *the restriction of $\succcurlyeq_1$ to $\mathcal{F}$ is complete.*

**Lemma 4.**

*Let $\succeq$ be an SPO with a representation $\mathcal{F}$. Set $\mathrm{Q}'_2 := \{-1_t + a1_\tau, a > 0, 0 \leq t < \tau\}$ and $\mathrm{Q}''_2 := \{-1_t + a1_\tau, a \geq 1, 0 \leq t < \tau\}$. The following conditions are equivalent:*

(a) $\succeq$ is $\mathrm{Q}_2$-complete;

(b) $\succeq$ is $\mathrm{Q}'_2$-complete;

(c) *the restriction of $\succcurlyeq_2$ to $\mathcal{F}$ is complete.*

*If every NPV functional from $\mathcal{F}$ has a positive discount function, then (a)–(c) are also equivalent to*

(d) $\succeq$ is $\mathrm{Q}''_2$-complete.

Lemma 3 (resp. Lemma 4) states that an SPO with a representation $\mathcal{F}$ is $\mathrm{Q}_1$-complete (resp. $\mathrm{Q}_2$-complete) if and only if for any $F, G \in \mathcal{F}$, either $\alpha \leq \beta$ or $\alpha \geq \beta$ (resp. $\mathrm{supp}\{\alpha\} = \mathrm{supp}\{\beta\}$ and the function $t \mapsto \alpha(t)/\beta(t)$ defined on $\mathrm{supp}\{\beta\}$ is monotone), where $\alpha$ and $\beta$ are the discount functions associated with $F$ and $G$. It also shows that to check $\mathrm{Q}_1$-completeness (resp. $\mathrm{Q}_2$-completeness) it is sufficient to test it on the set of projects $\mathrm{Q}'_1$ (resp. $\mathrm{Q}'_2$) with only two transactions. By characterizing $\mathrm{Q}_1$- and $\mathrm{Q}_2$-complete SPOs, Lemmas 3 and 4 solve problem A for $\mathrm{Q} = \mathrm{Q}_1, \mathrm{Q}_2$ (inasmuch as a function on $\mathrm{Q} \subseteq \mathrm{P}$ is a profitability metric if and only if it is a utility representation of the restriction of a $\mathrm{Q}$-complete SPO to $\mathrm{Q}$).

To illustrate, say, Lemma 4, consider the SPO $\succeq$ induced by the family of exponential discount functions, $\{t \mapsto e^{-\lambda t}, \lambda \in \mathrm{R}_+\}$. Clearly, the restriction of $\succeq_2$ to this family is complete. Thus, by Lemma 4, the SPO is $Q_2$-complete; moreover, it can be shown that IRR is a utility representation of the restriction of the SPO to $Q_2$ (i.e., IRR is a profitability metric on $Q_2$ induced by $\succeq$). This is a reformulation of a well-known fact that a project whose net cash flow has one change of sign possesses the IRR. Lemma 4 extends this result by describing all $Q_2$-complete SPO.

In order to describe $Q_3$- and $Q_4$-complete SPOs, we introduce the following notation. For $\alpha \in \mathcal{A} \setminus \{\chi\}$, denote by $H_\gamma^{(\alpha)}$, $\gamma \in [0, 1/\alpha(0+)]$ the NPV functional induced by the discount function $\gamma\alpha + (1-\gamma)\chi$. Set $H_\gamma^{(\chi)} := F^{(\chi)}$ for all $\gamma \in \mathrm{R}_+$. Note that $H_\gamma^{(\alpha)}(x) = x(0) + \gamma(F^{(\alpha)}(x) - x(0))$.

**Lemma 5.**

*Let $\succeq$ be an SPO with a representation $\mathcal{F}$. Set $Q_3' := \{-1_0 + a1_t + b1_\tau, (a,b,t,\tau) \in \mathrm{R}^2 \times \mathrm{R}_{++}^2\}$, $Q_4' := \{-1_s + a1_{s+t} + b1_{s+\tau}, (a,b,s,t,\tau) \in \mathrm{R}^2 \times \mathrm{R}_+ \times \mathrm{R}_{++}^2\}$, and $Q_4'' := \{x \in \mathrm{P} : x(0) \leq 0\}$. The following conditions are equivalent:*

(a) $\succeq$ *is $Q_3$-complete (resp. $Q_4$-complete or $Q_4''$-complete);*
(b) $\succeq$ *is $Q_3'$-complete (resp. $Q_4'$-complete);*
(c) *there are $\alpha \in \mathcal{A}$ and $\Gamma \subseteq [0,1]$ (resp. $\Gamma \subseteq (0,1]$) such that $\mathcal{F} = \{H_\gamma^{(\alpha)}, \gamma \in \Gamma\}$.*

To illustrate Lemma 5, consider again the SPO induced by the family of exponential discount functions $\{t \mapsto e^{-\lambda t}, \lambda \in \mathrm{R}_+\}$. It follows from Lemma 5 that the SPO is not $Q_3'$-complete. This is a reformulation of a well-known fact that an investment project with at least three transactions may have no IRR. One can use Lemma 5 to suggest a relevant profitability measure for projects from $Q_3'$ (or, equivalently, $Q_3$). Namely, from Lemma 5 it follows that if an SPO $\succeq$ is $Q_3$-complete, then there is $F \in \mathcal{NPV}$ such that for any $x, y \in Q_3$, $PI^F(x) \geq PI^F(y) \Rightarrow x \succeq y$, where $PI^F$ is the *profitability index* defined by $PI^F(x) := (F(x) - x(0))/(-x(0))$. This suggests the profitability index as a natural profitability measure for projects from $Q_3$.

### 4. Particular profitability metrics

The capital budgeting literature provides a variety of measures for project evaluation. In this section, we show that at least four of them – IRR, PP, DPP, and PI – are profitability metrics and find the corresponding consistent SPOs. This result is useful in at least three respects. First, it shows that these seemingly different metrics enjoy a unified interpretation as indicators of a project's financial stability across various economic scenarios. Second, it helps to characterize the largest sets of projects to which the metrics can be unequivocally extended. Third, the structure of the corresponding consistent SPOs suggests an appropriate situation for the use of each of these metrics.

Recall that the intersection of a collection of POs (SPOs) is a PO (SPO) (Lemma 1). Therefore, for any profitability metric $M$, the set of all $M$-consistent SPOs (treated as subsets of $\mathrm{P} \times \mathrm{P}$) contains the least element. This element describes the unambiguous part, which agrees with every $M$-consistent SPO. Given a profitability metric $M:Q \to R$, the greatest (with respect to inclusion) set $D \supseteq Q$ such that the least $M$-consistent SPO is $D$-complete is said to be the *natural domain of $M$*. The restriction of the least $M$-consistent SPO to the natural domain is said to be the *natural extension of $M$*. In what follows, a utility representation of the natural extension (if any) is also, rather loosely, referred to as the natural extension of $M$. The natural domain determines to what extent the profitability metric can be uniquely extended, and this unique extension is referred to as the natural one.

To motivate the idea behind the introduced concepts, consider the IRR capital budgeting technique. (The continuous version of) IRR is known as a genuine extension of the logarithmic rate of return, the metric defined over the set $Q_2''$ (where $Q_2''$ is defined in Lemma 4) of investment operations with only two transactions – an initial outlay and a final inflow – that sends each project $-1_t + b1_{t+\tau} \in Q_2''$ to its yield rate $(1/\tau)\ln b$. The questions of interest are: how is the IRR defined outside $Q_2''$ (note that the capital budgeting literature provides a variety of nonequivalent measures that reduce to the logarithmic rate of return for projects from $Q_2''$) and what is the largest set of projects on which the IRR is unambiguously defined? If the logarithmic rate of return (considered as a real-valued function on $Q_2''$) is a profitability metric, then the natural domain and the natural extension answer the questions. As another illustration, consider the PP capital budgeting technique. It is unambiguously defined on the set of projects, which we shall, for a moment, denote by $Q$, whose cumulative cash flows have one change of sign, from negative to nonnegative values. The definition of PP for a project whose cumulative cash flow has multiple changes of sign is more challenging (e.g., see Hajdasiński, 1993). The questions of interest are: can PP, defined on $Q$, be extended to a larger set and if it can, then how is the extension defined? Again, if PP is a profitability metric, then the natural domain and the natural extension answer the questions.

The natural domain need not exist. For instance, a constant function $M$ on $P_+$ is a profitability metric with the least $M$-consistent SPO induced by $\mathcal{NPV}$. For every $x \notin P_+ \cup (-P_{++})$ and $D \supseteq P_+ \cup \{x, -x\}$, this SPO is $P_+ \cup \{x\}$- and $P_+ \cup \{-x\}$-complete, but not $D$-complete as $x$ and $-x$ are incomparable by property 5 in Lemma 1. Thus, $M$ has no natural domain. The next lemma provides a sufficient condition for the existence of the natural domain and a way to construct it.

**Lemma 6.**

Given a profitability metric $M:Q \to R$, $Q \subseteq P$, set $\mathcal{F} = \bigcap \{F \in \mathcal{NPV}: I_{\{F\}^\circ}(x) \geq I_{\{F\}^\circ}(y)\}$, where the intersection is taken over all pairs $x, y \in Q$ such that $M(x) \geq M(y)$. The following statements hold.
(a) $\mathcal{F}$ is the representation of the least $M$-consistent SPO.
(b) If the preorder $\succcurlyeq$ on $\mathcal{F}$ induced by $Q$ is antisymmetric (i.e., $F \succcurlyeq G$ & $G \succcurlyeq F \Rightarrow F = G$), then the natural domain $D$ of $M$ exists and admits the representation
$$D = \bigcap \{x \in P : I_{\{F\}^\circ}(x) \geq I_{\{G\}^\circ}(x)\}, \tag{5}$$
where the intersection is taken over all pairs $F, G \in \mathcal{F}$ such that $F \succcurlyeq G$.

**Example 3.**

To illustrate the introduced concepts, consider the function $\pi : Q'_1 \to [1, +\infty)$ (where $Q'_1$ is defined in Lemma 3) given by $\pi(-1_0 + b1_\tau) := b$. $\pi - 1$ is just the undiscounted net benefit of a project. One can show (see Lemma 11 in the Appendix) that $\pi$ is a profitability metric whose natural extension is the undiscounted profitability index. To be more precise, $\pi$-consistent SPOs are induced by $\{H_\gamma^{(1_0)}, \gamma \in \Gamma\}$, where $(0,1) \subseteq \Gamma \subseteq [0,1]$. The representation of the least $\pi$-consistent SPO is $\{H_\gamma^{(1_0)}, \gamma \in [0,1]\}$. The natural domain of $\pi$ is $D = P \setminus \{x \in P : x(0) \geq 0, x(+\infty) < 0\}$ and the natural extension of $\pi$ is the complete preorder on $D$ with a utility representation $\bar{\pi} : D \to \bar{R}_+$ given by

$$\bar{\pi}(x) := \begin{cases} 0 & \text{if } x(0) < 0 \text{ and } x(+\infty) < 0 \\ (x(+\infty) - x(0))/(-x(0)) & \text{if } x(0) < 0 \text{ and } x(+\infty) \geq 0 \\ +\infty & \text{if } x(0) \geq 0 \text{ and } x(+\infty) \geq 0 \end{cases} \quad (6)$$

Note that if $x(0) < 0$ and $x(+\infty) \geq 0$, then $\bar{\pi}(x) = PI^F(x)$, where $F$ is the NPV functional induced by the discount function $1_0$. $\bar{\pi}$ genuinely extends $\pi$: $\bar{\pi}$ has the same natural properties as $\pi$ but significantly enlarges the set of comparable projects (from $Q'_1$ to $D$).

The capital budgeting literature knows several performance measures that reduce to the undiscounted net benefit $\pi$ for projects from $Q'_1$: the undiscounted profitability index and the undiscounted benefit-cost ratio (i.e., the ratio of the total cash benefits to the total cash costs), to mention just a few. The result above shows that the undiscounted profitability index is the only profitability metric among them. A similar conclusion can be derived for the discounted profitability index.

We proceed by showing that IRR, PP/DPP (more accurately, an order-reversing transformation of PP/DPP), PI as well as several other ratio type indices are profitability metrics. We describe their natural domains and natural extensions.

### 4.1. IRR

The purpose of this section is threefold. First, we introduce a generalization of the conventional definition of IRR to nonexponential families of discount functions. Second, we show that IRR (as well as the generalization) is a profitability metric, find its natural domain and natural extension, describe IRR-consistent SPOs, and provide their axiomatic characterization. Third, we show that (the continuous version of) the conventional IRR is a unique profitability metric whose restriction to $Q''_2$ (where $Q''_2$ is defined in Lemma 4) is the logarithmic rate of return, i.e., the metric that sends each project $-1_t + b1_{t+\tau} \in Q''_2$ to its yield rate, $(1/\tau) \ln b$.

The concept of internal rate of return does not necessary assume exponential discounting and can formally be applied to an arbitrary parametric family of discount functions indexed by a parameter interpreted as the discount rate. For instance, assuming power discounting (Harvey, 1986), one can define the internal rate of return of a project $x$ as the value of the discount rate $\lambda \in R_+$ under which $F_\lambda(x) = 0$, where $F_\lambda$ is the NPV functional associated with the discount function $t \mapsto (1+t)^{-\lambda}$. The following definition introduces a generic parametric family of discount functions that produces a consistent concept of IRR.

An indexed family $A := \langle \alpha_\lambda, \lambda \in R_+ \rangle$ of discount functions is said to be a *D-family* if the following two conditions hold: (I) each $\alpha_\lambda \in A$ is positive; (II) for any $0 \leq t < \tau$, the function $\lambda \mapsto \alpha_\lambda(\tau)/\alpha_\lambda(t)$ is strictly decreasing and onto $(0,1]$. In what follows, the NPV functional associated with a discount function $\alpha_\lambda \in A$ is denoted by $F_\lambda^{(A)}$. Set $\mathcal{F}^{(A)} := \{F_\lambda^{(A)}, \lambda \in R_+\}$. Condition (II) allows us to interpret parameter $\lambda$ as the degree of impatience.[11] Indeed, in the most general sense, one can define the degree of impatience as a characteristic of time preference that, when increased, makes the earlier of any two timed outcomes more preferable. This is exactly what the strict decreasingness of $\lambda \mapsto \alpha_\lambda(\tau)/\alpha_\lambda(t)$ asserts: for any $t < \tau$ and $a, b \in R_{++}$, $F_\lambda^{(A)}(a1_t) = F_\lambda^{(A)}(b1_\tau) \Rightarrow F_{\lambda'}^{(A)}(a1_t) > F_{\lambda'}^{(A)}(b1_\tau) \quad \forall \lambda' > \lambda$. Put differently, if $\lambda' > \lambda$, then $\alpha_{\lambda'}$ exhibits greater impatience than $\alpha_\lambda$ in the sense of Quah and Strulovici (2013, Section II.C). Provided that elements of $A$ are differentiable, condition (II) implies that for any $t$, the instantaneous discount rate, $-(\ln \alpha_\lambda(t))'$, is a nondecreasing function of $\lambda$. In the special case when the function $(\lambda, t) \mapsto \alpha_\lambda(t)$ is continuously differentiable an analogue of a D-family governed by a real (rather than nonnegative real) parameter $\lambda$ and its relation to the concept of IRR is studied in Vilensky and Smolyak (1999).

Given a strictly decreasing discount function $\alpha$, the family $\langle \alpha^\lambda, \lambda \in R_+ \rangle$, called a *power family*, serves as an example of a D-family. In particular, the exponential discounting family $E := \langle t \mapsto e^{-\lambda t}, \lambda \in R_+ \rangle$, constant sensitivity discounting families $\langle t \mapsto \exp(-\lambda t^\beta), \lambda \in R_+ \rangle$, $\beta > 0$ (Ebert and Prelec, 2007), and generalized hyperbolic discounting families $\langle t \mapsto (1+\beta t)^{-\lambda/\beta}, \lambda \in R_+ \rangle$, $\beta > 0$ (Loewenstein and Prelec, 1992; Weitzman, 2001) are power and, hence, D-families.

Given a D-family $A$, it follows from a convergence theorem (Monteiro et al., 2018, Theorem 6.8.6, p. 188) that for any $x \in P$, the function $g_x^{(A)}(\lambda) := F_\lambda^{(A)}(x)$, defined on $R_+$, is continuous. If it has one change of sign, the internal rate of return is defined as follows. A project $x$ is said to *possess the IRR w.r.t.* $A$ if there exists a number $IRR^{(A)}(x) \in R_+$ such that $\operatorname{sgn} g_x^{(A)}(\lambda) = \operatorname{sgn}(IRR^{(A)}(x) - \lambda)$ for all $\lambda \in R_+$. Put differently, $x$ possesses the IRR w.r.t. $A$ if $g_x^{(A)}$ has a unique root, and at this root, the function changes sign from positive to negative. If $A = E$, this definition reduces to the conventional definition of IRR. Denote by $Q^{(A)} \subset P$ the set of projects possessing the IRR w.r.t. $A$.

Clearly, $Q_2'' \subset Q^{(A)}$ for any D-family $A$. The restriction of $IRR^{(A)}: Q^{(A)} \to R_+$ to $Q_2''$, denoted by $RR^{(A)}$, is called the *rate of return w.r.t.* $A$. $RR^{(A)}$ sends each project $-1_t + b1_\tau \in Q_2''$ to the solution $\lambda \in R_+$ of the equation $\alpha_\lambda(t) = b\alpha_\lambda(\tau)$. For instance, if $A$ is a power family $\langle \alpha^\lambda, \lambda \in R_+ \rangle$, then $RR^{(A)}(-1_t + b1_\tau) = (\ln \alpha(t) - \ln \alpha(\tau))^{-1} \ln b$.

Our next result shows that $RR^{(A)}$ and $IRR^{(A)}$ are profitability metrics.

**Proposition 3.**

---

[11] Given that cash flows in our model are deterministic, discounting (which normally encompasses both the impatience of the investor – risk-free rate – and a risk premium) reflects impatience only.

*Let* A *be a D-family. The following statements hold.*

(a) $RR^{(A)}$ *is a profitability metric.*

(b) *An SPO is $RR^{(A)}$-consistent if and only if it is induced by $\{F_\lambda^{(A)}, \lambda \in \Lambda\}$, where $\Lambda$ is a dense subset of $R_+$.*

(c) *The least $RR^{(A)}$-consistent SPO is induced by $\mathcal{F}^{(A)}$.*

(d) *The natural domain of $RR^{(A)}$, denoted by $D^{(A)}$, consists of projects $x \in P$ such that the function $g_x^{(A)}$ is either nonnegative or negative, or there is $\lambda \in R_+$ such that $g_x^{(A)}$ is nonnegative on $[0, \lambda]$ and negative on $(\lambda, +\infty)$.*

(e) *The natural extension of $RR^{(A)}$ is the complete preorder on $D^{(A)}$ with a utility representation $\overline{RR}^{(A)} : D^{(A)} \to \bar{R}$ given by $\overline{RR}^{(A)}(x) := \sup\{\lambda \in R_+ : g_x^{(A)}(\lambda) \geq 0\}$ (with the convention $\sup \varnothing = -\infty$).*

*Moreover, statements (a)–(e) remain valid with $RR^{(A)}$ replaced by $IRR^{(A)}$.*

Proposition 3 demonstrates that the rate of return w.r.t. A defined for investment operations with only two transactions $Q_2''$ admits a unique extension (satisfying several reasonable conditions) to $Q^{(A)}$. This extension is exactly $IRR^{(A)}$. The literature knows several performance measures that reduce to the logarithmic rate of return $RR^{(E)}$ (or an order-preserving transformation of this value) being restricted to $Q_2''$: the conventional IRR, the modified IRR (Lin, 1976), and the metrics introduced in Arrow and Levhari (1969) and Bronshtein and Skotnikov (2007), to mention just a few. Proposition 3 shows that the conventional IRR is the only profitability metric among them.

If the function $g_x^{(E)}$ has multiple roots, the literature suggests various modifications of IRR that reduce to the conventional IRR whenever $g_x^{(E)}$ has one change of sign. For instance, the minimal root is important as the asymptotic growth rate of a sequence of repeated projects (Cantor and Lippman, 1983). In contrast, Bidard (1999) advocated the maximal root. More involved selection procedures among the roots were proposed in Hartman and Schafrick (2004) and Weber (2014).[12] A variety of completely different modifications of IRR were introduced in Promislow and Spring (1996). Proposition 3 shows that these modifications are not profitability metrics.

As established in Proposition 3, the natural extension of $IRR^{(E)}$ denoted by $\overline{RR}^{(E)}$ sends $x \in D^{(E)}$ to $-\infty$ if $g_x^{(E)}$ is negative, to $+\infty$ if $g_x^{(E)}$ is nonnegative, and to the value $\lambda \in R_+$ if $g_x^{(E)}$ is nonnegative on $[0, \lambda]$ and negative on $(\lambda, +\infty)$. Let us show that $\overline{RR}^{(E)}$ also genuinely extends $IRR^{(E)}$ as an accept/reject investment decision-making tool (recall that $IRR^{(E)}$ provides such a tool when compared with a hurdle rate). It is widely accepted in the capital budgeting literature that any such tool should be compatible with the NPV criterion (e.g., see Hartman and Schafrick, 2004; Marchioni and Magni, 2018). We shall say that a function $M : Q \to \bar{R}$ (interpreted as a measure of yield) defined on a nonempty set $Q \subseteq P$ is *NPV-compatible* if for any $x \in Q$ and $\lambda \in R_+$, $M(x) \geq \lambda \iff F_\lambda^{(E)}(x) \geq 0$. In words, $M$ is NPV-compatible if and only if for any discount rate $\lambda \in R_+$, it supplies the same recommendation in accept/reject decisions as the NPV criterion. Note that the negative part of an NPV-compatible function $M$ is immaterial in the sense that $M$ and $x \mapsto \begin{cases} M(x) \text{ if } M(x) \geq 0 \\ -\infty \text{ otherwise} \end{cases}$ provide the same accept/reject decisions. So we may restrict the codomain of an NPV-compatible function to $\{-\infty\} \cup \overline{R}_+$. The conventional rate of return and IRR – $RR^{(E)}$ and $IRR^{(E)}$ – are well-known examples of NPV-compatible measures of yield. Just from the definition of NPV-compatibility, we get that $\overline{RR}^{(E)}$ is the most general NPV-compatible measure of yield. Namely, a function $M : Q \to \{-\infty\} \cup \overline{R}_+$ defined on a nonempty set $Q \subseteq P$ is NPV-compatible if and only if $Q \subseteq D^{(E)}$ and $M$ is the restriction of $\overline{RR}^{(E)}$ to $Q$.

---

[12] We also refer to Hazen (2003) for an interesting result that the choice of a particular root is in some sense immaterial and Osborne (2010), who argued to take into account all the roots of the IRR polynomial, real and complex.

Note that from an economic viewpoint, $D^{(E)}$ adds almost nothing to $Q^{(E)}$. Roughly speaking, $IRR^{(E)}$ (more generally, $IRR^{(A)}$) does not possess an extension to a larger set.[13] The picture changes if we allow incomparable pairs of projects. Parts (b) and (c) of Proposition 3 describe $IRR^{(A)}$-consistent SPOs and their common part. Enjoying the same natural properties as $IRR^{(A)}$, they substantially enlarge the set of comparable projects, as is illustrated in the following example. Note that the fact that the SPO with the representation $\mathcal{F}^{(E)}$ can be considered as an extension of $IRR^{(E)}$ was observed in Bronshtein and Akhmetova (2004).[14]

**Example 4.**
1. Consider projects $x = 1_0 - 100 \cdot 1_1 + 120 \cdot 1_2$, $y = -100 \cdot 1_1 + 120 \cdot 1_2$, and $z = -100 \cdot 1_1 + 230 \cdot 1_2 - 132 \cdot 1_3$. Project $y$ can be interpreted as the lender's cash flow for a loan at 20% per period. Project $x$ is the same loan accompanied with an application fee of 1% of the loan amount paid one period prior. Project $z$ can be interpreted as loan $y$, in which the borrower repays by mistake 230 (rather than 120) money units at time 2, and the difference is refunded at time 3 with 20% interest. Note that $x$ and $z$ are examples of so-called mixture projects, i.e., are neither pure investments nor pure borrowings. As a result, $x$, $y$, and $z$ are incomparable with respect to the conventional IRR: the IRR equation for $x$ (i.e., $g_x^{(E)}(\lambda) = 0$) has the roots 0.19 and 4.59, the IRR equation for $z$ has the roots 0.1 and 0.18, whereas $y$ possesses the IRR 0.18. At least for $x$ and $y$, this incomparability is counterintuitive as $x$ is more favorable to the lender than $y$. In contrast, we have $x \succ y \succ z$ for any $IRR^{(E)}$-consistent SPO $\succeq$ as $\{\lambda \in R_+ : g_x^{(E)}(\lambda) \geq 0\} = [0, 0.19] \cup [4.59, +\infty)$, $\{\lambda \in R_+ : g_y^{(E)}(\lambda) \geq 0\} = [0, 0.18]$, $\{\lambda \in R_+ : g_z^{(E)}(\lambda) \geq 0\} = [0.1, 0.18]$, and $[0, 0.19] \cup [4.59, +\infty) \supset [0, 0.18] \supset [0.1, 0.18]$. That is, the set of discount rates for which the project is profitable (i.e., has nonnegative NPV) is larger for $x$ than for $y$, and larger for $y$ than for $z$.

---

[13] We want to stress that we treat conditions NP, MON, INT, and USC as minimally reasonable for a binary relation on $P$ to be relevant for decision making. Thus, part (e) of Proposition 3 is an argument against various generalizations of the conventional definition of IRR. However, this result does not assert to use the conventional definition of IRR instead of its restriction to some proper subset. In particular, it does not contradict Herbst (1978), Gronchi (1986), and Promislow (2015, Section 2.12), who provide arguments that IRR is meaningful for conventional or so-called pure investments only.

[14] Being applied to valuation of cash flow (or, more generally, utility) streams rather than profitability, multiple discount rates have gained increasing popularity in the recent literature (Chambers and Echenique, 2018; Drugeon et al., 2019).

2. A conventional tool in accept/reject investment decision making is based on the comparison of the IRR with a hurdle rate. An $IRR^{(E)}$-consistent SPO $\succeq$ extends this tool in two directions. First, it enlarges the set of accepted and rejected projects. Consider again projects $x$, $y$, and $z$ from part 1 and assume that the hurdle rate is 20%. Since the hurdle rate can be earned by project $y$ (instead of $y$ we can take any project with the IRR 20%) and $x \succ y \succ z$, project $x$ should be accepted, and project $z$ should be rejected. Recall that $x$ and $z$ have no IRR, so the accept/reject decision making based on the conventional IRR is inapplicable in this case. Second, an $IRR^{(E)}$-consistent SPO extends the notion of opportunity cost by considering investment alternatives that have no IRR. For instance, consider a microfinance institution whose typical product is a loan at 20% per period. Then, when analyzing a new investment opportunity, a reasonable choice of the institution's hurdle rate would be 20%. Now assume that the typical product is a loan at 20% per period accompanied with an application fee of 1% of the loan amount paid one period prior. The institution's cash flow for this loan is $cx$, where $c > 0$ is the loan amount measured in hundreds of money units. Given that $x$ has no IRR, what is the institution's hurdle rate? An $IRR^{(E)}$-consistent SPO $\succeq$ avoids this question, by comparing an investment opportunity with $cx$ directly. For instance, investment opportunities $y$ and $z$ should be rejected as $cx \sim x \succ y \succ z$.

3. Consider a line of credit consisting of a collection of loans $x_1,...,x_n$ at variable (logarithmic) interest rates of, respectively, $r_1,...,r_n$, where, without loss of generality, we assume that $r_1 \geq ... \geq r_n$. One expects that the effective interest rate of the whole line of credit $x_1 + ... + x_n$ is somewhere between $r_1$ and $r_n$. And indeed this is the case as $IRR^{(E)}(x_1) = r_1$ & ... & $IRR^{(E)}(x_n) = r_n \Rightarrow r_1 = \max\{r_1,...,r_n\} \geq IRR^{(E)}(x_1 + ... + x_n) \geq \min\{r_1,...,r_n\} = r_n$, provided that $x_1 + ... + x_n$ has the IRR. The problem is that, in general, $x_1 + ... + x_n$ has no IRR (the case $r_1 = r_n$ is a notable exception). In contrast, as expected, we have $x_1 \succeq x_1 + ... + x_n \succeq x_n$ for any $IRR^{(E)}$-consistent SPO $\succeq$; moreover, the inequalities are strict whenever $r_1 > r_n$. Furthermore, the inequalities $x_1 \succeq x_1 + ... + x_n \succeq x_n$ remain valid for any projects $x_1,...,x_n$ with $x_1 \succeq ... \succeq x_n$ regardless of whether they have or have no IRR. By way of illustration, consider projects $x_1 = -1_0 + 19 \cdot 1_1$ and $x_2 = -100 \cdot 1_2 + 150 \cdot 1_3$. These are conventional loans (from the lender's perspective) with the logarithmic interest rates $IRR^{(E)}(x_1) = \ln(19/1) \approx 2.94$ and $IRR^{(E)}(x_2) = \ln(150/100) \approx 0.41$. Project $x_1 + x_2$ is a line of credit with (dramatically) variable logarithmic interest rate, from 2.94 in the first period to 0.41 in the third period. The line of credit $x_1 + x_2$ is incomparable with $x_1$ and $x_2$ in the sense of IRR: the IRR equation for $x_1 + x_2$ has three roots, 0.97, 1.61, and 2.43. In contrast, as expected, we have $x_1 \succ x_1 + x_2 \succ x_2$ for any $IRR^{(E)}$-consistent SPO $\succeq$ as $\{\lambda \in R_+ : g^{(E)}_{x_1}(\lambda) \geq 0\} = [0, 2.94]$, $\{\lambda \in R_+ : g^{(E)}_{x_1+x_2}(\lambda) \geq 0\} = [0, 0.97] \cup [1.61, 2.43]$, $\{\lambda \in R_+ : g^{(E)}_{x_2}(\lambda) \geq 0\} = [0, 0.41]$, and $[0, 2.94] \supset [0, 0.97] \cup [1.61, 2.43] \supset [0, 0.41]$.

4. The modified IRR (MIRR) is a popular performance measure designed to eliminate the issue of multiple IRRs. It is defined as the IRR of a modified cash flow and accounts for financing and reinvestment rates. Although MIRR is an umbrella term for a variety of performance measures, all of them (in their continuous version) reduce to $RR^{(E)}$ on $Q''_2$. For each fixed financing and reinvestment rates (and a way to modify a cash flow), MIRR is not identically equal to $IRR^{(E)}$ on $Q^{(E)}$ and, therefore, by Proposition 3, is not a profitability metric. Let us illustrate this point for a version of MIRR defined in Lin (1976) and Beaves (1988). These authors define MIRR as the IRR of the modified cash flow stream obtained by discounting (at the financing rate) all negative cash flows to time 0 and compounding (at the reinvestment rate) all positive cash flows to the terminal time. Assume that the financing and reinvestment rates are both 10% and consider projects $u = -100 \cdot 1_0 + 105 \cdot 1_1 + 25 \cdot 1_2$ and $v = -100 \cdot 1_0 - 90 \cdot 1_1 + 275 \cdot 1_2$. Then, the MIRRs of $u$, $v$, and $u+v$ are $((105 \cdot 1.1 + 25)/100)^{(1/3)} - 1 = 19\%$, $(275/(100 + 90/1.1))^{(1/3)} - 1 = 23\%$, and $((15 \cdot 1.1 + 300)/200)^{(1/3)} - 1 = 26\%$, respectively. This demonstrates that MIRR is not a profitability metric (if it was, the MIRR of $u+v$ would be somewhere between the MIRRs of $u$ and $v$).

Investment and financing projects require formally distinct definitions of IRR. For financing projects, the conventional IRR is defined if the IRR polynomial has a unique root, and at this root, the polynomial changes sign from negative to positive. For financing projects, the greater the IRR, the less favorable the project. Thus, prior to use the IRR capital budgeting technique, a manager has to determine the investment/financing status of a project. An $IRR^{(E)}$-consistent SPO generalizes both the definitions of IRR and, therefore, does not require to determine the investment/financing status of a project. Indeed, it follows from the definition of the SPO that for a project from $Q^{(E)}$ (resp. $-Q^{(E)}$), the greater the IRR, the more (resp. less) profitable the project.

The least $IRR^{(E)}$-consistent SPO is related to a modification of IRR introduced in Promislow and Spring (1996). According to one of the constructs proposed by these authors, a project $x$ is more profitable than $y$ if the set $\{\lambda : g_x^{(E)}(\lambda) \geq 0\}$ has larger Lebesgue measure than $\{\lambda : g_y^{(E)}(\lambda) \geq 0\}$, whereas, according to the least $IRR^{(E)}$-consistent SPO $\succeq$, $x \succeq y$ if the former set is a superset of the latter.

In order to extend the class of projects possessing the IRR, some authors argue (e.g., see Vilensky and Smolyak, 1999) to take into account roots of $g_x^{(E)}$ only in a reasonable range $[\lambda_-, \lambda_+]$, where $\lambda_-$ and $\lambda_+$ are the least and greatest feasible interest rates. Clearly, this modification is a profitability metric; a corresponding consistent SPO is induced by $\{F_\lambda^{(E)}, \lambda \in [\lambda_-, \lambda_+]\}$.

In order to formulate our next result we introduce the following definition. Set $S := R_{++} Q_2'' = \{-a 1_t + b 1_\tau, 0 < a \leq b, 0 \leq t < \tau\}$; to simplify the notation we write $(a,b;t,\tau)$ for $-a 1_t + b 1_\tau \in S$. A function $M : S \to R$ is said to be a *rate of return* if the following four conditions hold.
1. $M(a,b;t,\tau) = M(\lambda a, \lambda b; t, \tau)$ for any $\lambda > 0$.
2. $M(a,b;t,\tau) = M(b,c;\tau,\delta) \Rightarrow M(a,b;t,\tau) = M(a,c;t,\delta)$.
3. $M$ is strictly increasing in its second argument.
4. For any $x \in S$ and $0 \leq t < \tau$, there are $0 < a \leq b$ such that $M(a,b;t,\tau) = M(x)$.

The value $M(a,b;t,\tau)$ is interpreted as the yield rate (the rate of return) of the project $(a,b;t,\tau)$. Condition 1 states that a rate of return takes no account of the investment size and hence is a relative measure. According to condition 2, if rates of return over two subsequent periods are equal, the rate of return over the consolidated period will be the same. By condition 3, a rate of return is an increasing function of the final inflow. Finally, according to condition 4, delay can always be compensated by changing a money flow. An example of a rate of return is provided by the logarithmic rate of return, $M(a,b;t,\tau) = (\tau-t)^{-1}\ln(b/a)$.

Although the concept of IRR w.r.t. a D-family seems to be intuitive, it is introduced without justification, just by analogy with the conventional IRR. Our next result shows that it is actually a genuine extension of a rate of return. Namely, the IRR w.r.t. a D-family is, up to an order-preserving transformation, the only profitability metric whose restriction to $S$ is a rate of return.

**Proposition 4.**

*For an SPO $\succeq$, the following conditions are equivalent:*

(a) *there exists a D-family $A$ such that $\succeq$ is $IRR^{(A)}$-consistent;*

(b) *there exists a rate of return $M$ such that $\succeq$ is $M$-consistent;*

(c) *the restriction of $\succeq$ to $Q_2''$ is a lower semicontinuous (in the subspace topology) complete preorder and $-1_t + a1_\tau \succ -1_t + b1_\tau$, whenever $1 \leq b < a$ and $0 \leq t < \tau$.*

It follows from Proposition 4 that the IRR w.r.t. a D-family is, up to an order-preserving transformation, the only profitability metric whose restriction to $Q_2''$ satisfies the two natural conditions – continuity and monotonicity. Another interesting consequence of Proposition 4 is that a rate of return is nondecreasing in its third argument and nonincreasing in the fourth argument, that is, delay is undesirable. This follows from the definition of $RR^{(A)}$ and nonincreasingness of a discount function. Note that conditions 1–4 do not contain any explicit assumption on how a rate of return depends on time.

We proceed by characterizing SPOs consistent with the conventional IRR. For any $x \in P$ and $\tau > 0$, put $x^{(+\tau)}(t) := \begin{cases} 0 & \text{if } t < \tau \\ x(t-\tau) & \text{if } t \geq \tau \end{cases}$. That is, $x^{(+\tau)}$ is the project $x$ postponed until $\tau$. A binary relation $\succeq$ on a set $Q \subseteq P$ (with $\sim$ being the symmetric part of $\succeq$) is said to be *stationary* if $x \sim x^{(+\tau)}$, whenever $\tau > 0$ and $x, x^{(+\tau)} \in Q$. Stationarity states that postponement of a project does not affect profitability. Note that if a project $x \in P$ possesses the IRR w.r.t. $E$, then so does $x^{(+\tau)}$ and $IRR^{(E)}(x) = IRR^{(E)}(x^{(+\tau)})$. Our next result characterizes $IRR^{(E)}$-consistent SPOs by means of stationarity and monotonicity conditions.

**Proposition 5.**

*For an SPO $\succeq$, the following conditions are equivalent:*

(a) *$\succeq$ is $IRR^{(E)}$-consistent;*

(b) *the restriction of $\succeq$ to $Q_2''$ is stationary and there exists $t > 0$ such that $-1_0 + a1_t \succ -1_0 + b1_t$ for any $a > b \geq 1$.*

Proposition 5 shows that $IRR^{(E)}$ is, up to an order-preserving transformation, the only profitability metric whose restriction to $Q_2''$ is stationarity and monotone. This characterization could be predicted in view of Proposition 4 and a well-known fact that multiplicative discounting reduces to exponential discounting under stationarity (Fishburn and Rubinstein, 1982, Theorem 2).

By Proposition 3 (part (b)), there are $\beth_2$ (the cardinality of the set of all dense subsets of $\mathbb{R}_+$) $IRR^{(A)}$-consistent SPOs, each of which we consider as an extension of $IRR^{(A)}$. We proceed by showing that these extensions are essentially unique, namely, they coincide on a large class of projects, called regular. Given $\mathcal{F} \subseteq \mathcal{NPV}$, a project $x \in P$ is said to be *regular w.r.t.* $\mathcal{F}$ if $\{x\}^\circ \cap \mathcal{F}$ is a regular closed set (i.e., is equal to the closure of its interior) in the subspace topology on $\mathcal{F}$. Denote by $\mathcal{R}(\mathcal{F})$ the set of all projects that are regular w.r.t. $\mathcal{F}$. The following lemma motivates the definition.

**Lemma 7.**

*Let $\succeq$ be an SPO with a representation $\mathcal{F}$, $\mathcal{F}'$ be a dense subset of $\mathcal{F}$, and $\succeq'$ be the SPO induced by $\mathcal{F}'$. Then the restrictions of $\succeq$ and $\succeq'$ to $\mathcal{R}(\mathcal{F})$ coincide.*

To illustrate Lemma 7, assume that in the notation of the lemma $\mathcal{F}$ is open in the subspace topology on $\{F \in P^* : F(1_0) = 1\}$ and convex; then $\mathcal{R}(\mathcal{F}) = P$, so we have $\succeq = \succeq'$. This result can be established with the help of the fact that if a convex subset C of a topological vector space has a nonempty interior, then $\mathrm{cl}(C) = \mathrm{cl}(\mathrm{int}(C))$ (Aliprantis and Border, 2006, Lemma 5.28, p. 182). We omit the details.

Given a D-family A, it can be shown that the map $\lambda \mapsto F_\lambda^{(A)}$ is a homeomorphism between $\mathbb{R}_+$ and $\mathcal{F}^{(A)}$ endowed with the subspace topology (see Lemma 10 in the Appendix). Thus, Lemma 7 implies that the structure of the dense subset $\Lambda$ of $\mathbb{R}_+$ in part (b) of Proposition 3 is immaterial, provided that we restrict ourselves to regular projects w.r.t. $\mathcal{F}^{(A)}$. Moreover, a project $x \in P$ is regular w.r.t. $\mathcal{F}^{(A)}$ if and only if $A(x) := \{\lambda \in \mathbb{R}_+ : g_x^{(A)}(\lambda) \geq 0\}$ is regular closed in $\mathbb{R}_+$. By way of illustration, consider a project $x \in P$ such that the set of solutions $\{\lambda\}$ of the IRR equation, $g_x^{(A)}(\lambda) = 0$, is finite. For instance, this condition holds in the practically relevant case when A is a power family and $x$ is a nonzero finite project (Tossavainen, 2006). Since $g_x^{(A)}$ is continuous, $A(x)$ is a union of finitely many closed intervals. Therefore, $x$ is regular w.r.t. $\mathcal{F}^{(A)}$ if and only if $A(x)$ has no isolated points, that is, $g_x^{(A)}$ has no zero local maxima. The latter condition is rather weak; so, roughly speaking, all real-world projects are regular.

### 4.2. PP and DPP

In this section, we show that an order-reversing transformation of the payback period, as well as of its discounted counterpart, is a profitability metric. We find its natural domain, the natural extension, and describe corresponding consistent SPOs.

For any $x \in P$ and $\tau \in \mathbb{R}_+$, denote by $x_{\leq \tau}(t) := x(t)I_{[0,\tau]}(t) + x(\tau)I_{(\tau,+\infty)}(t)$ the project truncated at $\tau$. Note that $x_{\leq \tau} \in P$. Given $\alpha \in \mathcal{A}$, set

$$G_\tau^{(\alpha)}(x) := F^{(\alpha)}(x_{\leq \tau}) = x(0) + \int_0^\tau \alpha(t)\mathrm{d}x(t). \tag{7}$$

We have $G_\tau^{(\alpha)} \in \mathcal{NPV}$; indeed, from (7) it follows that $G_\tau^{(\alpha)}$ is the NPV functional induced by the discount function $\alpha I_{[0,\tau]}$. The function $\tau \mapsto G_\tau^{(\alpha)}(x)$ represents the cumulative discounted cash flow associated with $x$. If it has one change of sign over $R_{++}$, the discounted payback period is defined as follows. A project $x \in P$ is said to *possess the DPP w.r.t.* $\alpha$ if there exists a number $DPP^{(\alpha)}(x) \in R_{++}$ such that $G_\tau^{(\alpha)}(x) < 0$ (resp. $G_\tau^{(\alpha)}(x) \geq 0$) for any $\tau \in (0, DPP^{(\alpha)}(x))$ (resp. $\tau \in [DPP^{(\alpha)}(x), +\infty)$). In words, $DPP^{(\alpha)}(x)$ is the first positive date where the cumulative discounted cash flow associated with $x$ is nonnegative. The conventional definitions of PP and DPP correspond to $\alpha = 1_0$ and $\alpha(t) = e^{-\lambda t}$, $\lambda \in R_{++}$, respectively. Let $Q^{(\alpha)} \subset P$ be the set of projects possessing the DPP w.r.t. $\alpha$. Note that $Q^{(\alpha)} \neq \varnothing$ unless $\alpha = \chi$.

The reciprocals of PP and DPP are known to be crude estimates of the IRR (Gordon, 1955; Sarnat and Levy, 1969; Bhandari, 2009). Our next result shows that $RDPP^{(\alpha)}(x) := 1/DPP^{(\alpha)}(x)$ is a profitability metric and describes its natural domain, the natural extension, and $RDPP^{(\alpha)}$-consistent SPOs.

**Proposition 6.**

Let $\alpha \in \mathcal{A} \setminus \{\chi\}$, $Q_-^{(\alpha)} := \{x \in P: G_\tau^{(\alpha)}(x) < 0$ for all $\tau \in R_{++}\}$, and $Q_+^{(\alpha)} := \{x \in P: H_{1/\alpha(0+)}^{(\alpha)}(x) \geq 0$ and $G_\tau^{(\alpha)}(x) \geq 0$ for all $\tau \in R_{++}\}$.[15] The following statements hold.

(a) $RDPP^{(\alpha)} : Q^{(\alpha)} \to R_{++}$ is a profitability metric.

(b) An SPO is $RDPP^{(\alpha)}$-consistent if and only if it is induced by $\{G_\tau^{(\alpha)}, \tau \in T\} \cup \{H_\gamma^{(\alpha)}, \gamma \in \Gamma\}$, where T is a dense subset of $\text{int}(\text{supp}\{\alpha\})$ and $\Gamma \subseteq [1, 1/\alpha(0+)]$.

(c) The least $RDPP^{(\alpha)}$-consistent SPO is induced by $\{G_\tau^{(\alpha)}, \tau \in \text{int}(\text{supp}\{\alpha\})\} \cup \{H_\gamma^{(\alpha)}, \gamma \in [1, 1/\alpha(0+)]\}$.

(d) The natural domain of $RDPP^{(\alpha)}$ is $D^{(\alpha)} := Q_-^{(\alpha)} \cup Q^{(\alpha)} \cup Q_+^{(\alpha)}$.

(e) The natural extension of $RDPP^{(\alpha)}$ is the complete preorder on $D^{(\alpha)}$ with a utility representation $\overline{RDPP}^{(\alpha)} : D^{(\alpha)} \to \overline{R}$ given by

$$\overline{RDPP}^{(\alpha)}(x) := \begin{cases} \sup\{\gamma \in [\alpha(0+), 1] : H_{1/\gamma}^{(\alpha)}(x) \geq 0\} - 1 & \text{if } x \in Q_-^{(\alpha)} \\ RDPP^{(\alpha)}(x) & \text{if } x \in Q^{(\alpha)} \\ +\infty & \text{if } x \in Q_+^{(\alpha)} \end{cases} \quad (8)$$

*(with the convention* $\sup \varnothing = -\infty$*).*

In the case when the function $\tau \mapsto G_\tau^{(\alpha)}(x)$ has multiple changes of sign, some authors suggest to define DPP as the minimum time $t$ (if any) such that $G_\tau^{(\alpha)}(x) \geq 0$ for all $\tau \geq t$ (e.g., see Hajdasiński, 1993). Even though this definition seems to be intuitive from an economic viewpoint, Proposition 6 shows that an order-reversing transformation of the DPP defined this way is not a profitability metric and therefore, contrary to the claim (Hajdasiński, 1993, p. 184), is unable to serve for profitability measurement purposes. The largest extension of the domain of $DPP^{(\alpha)}$ is described in part (d). Unfortunately, from an economic viewpoint, $D^{(\alpha)}$ adds almost nothing to $Q^{(\alpha)}$. Loosely speaking, PP and DPP do not possess an extension to a larger set.

---

[15] Recall that $H_\gamma^{(\alpha)}$, $\alpha \in \mathcal{A}$, $\gamma \in R_+$ is the functional on P defined by $H_\gamma^{(\alpha)}(x) = x(0) + \gamma(F^{(\alpha)}(x) - x(0))$.

The situation changes if we allow incomparable pairs of projects. Parts (b) and (c) describe $RDPP^{(\alpha)}$-consistent SPOs and their common part. In the next example, their implications are compared with the implications of the conventional definition of DPP and the definition of DPP in Hajdasiński (1993).

**Example 5.**

Consider projects $x = 1_0 - 100 \cdot 1_1 + 120 \cdot 1_2$, $y = -50 \cdot 1_0 + 72 \cdot 1_2$, and $z = -255 \cdot 1_0 + 306 \cdot 1_1 - 60 \cdot 1_2 + 72 \cdot 1_3$, where time is measured in years. Project $x$ is borrowed from part 1 of Example 4. Project $y$ is the lender's cash flow for a 2-year loan at 20% per year. Project $z$ can be interpreted as the lender's cash flow for a line of credit consisting of two 1-year loans at 20% interest. These projects are incomparable with respect to PP: $y$ has PP 2 years, whereas $x$ and $z$ have no PP (i.e., they do not possess the DPP w.r.t. $1_0$). Additionally, note that $z$ (resp. $x$ and $y$) has PP 3 (resp. 2) years in the sense of Hajdasiński.

Let $\succeq$ be an $RDPP^{(1_0)}$-consistent SPO. From Proposition 6 it follows that $\succeq$ has a representation $\{G_\tau^{(1_0)}, \tau \in T\}$, where $T$ is a dense subset of $R_+$. Thus, for any $u, v \in P$, $u \succeq v$ if and only if $\{\tau \in T : G_\tau^{(1_0)}(u) \geq 0\} \supseteq \{\tau \in T : G_\tau^{(1_0)}(v) \geq 0\}$. If projects $u$ and $v$ are finite, then the latter inclusion holds if and only if $\{\tau \in R_+ : G_\tau^{(1_0)}(u) \geq 0\} \supseteq \{\tau \in R_+ : G_\tau^{(1_0)}(v) \geq 0\}$. Or, in words, $u \succeq v$ if and only if the set of dates $\tau$ for which the project truncated at $\tau$ has nonnegative total undiscounted value is larger for $u$ than for $v$. For projects $x$, $y$, and $z$, defined above, we have $\{\tau \in R_+ : G_\tau^{(1_0)}(x) \geq 0\} = [0,1) \cup [2, +\infty)$, $\{\tau \in R_+ : G_\tau^{(1_0)}(y) \geq 0\} = [2, +\infty)$, and $\{\tau \in R_+ : G_\tau^{(1_0)}(z) \geq 0\} = [1,2) \cup [3, +\infty)$. As $[0,1) \cup [2, +\infty) \supset [2, +\infty)$, we get $x \succ y$, whereas the remaining pairs of projects are incomparable.

Now assume that an investor decides to undertake projects $y$ and $z$. Since $y$ and $z$ have PP, respectively, 2 and 3 years in the sense of Hajdasiński, he/she expects that the PP of their union, $y + z$, is somewhere between 2 and 3. However, $y + z = -305 \cdot 1_0 + 306 \cdot 1_1 + 12 \cdot 1_2 + 72 \cdot 1_3$ has PP 1 year, which is counterintuitive. By definition, such type of inconsistency cannot occur with $\succeq$. For instance, using $\succeq$, in this example we also get $y + z \succ y$ and $y + z \succ z$. However, since $y$ and $z$ are incomparable with respect to $\succeq$, inconsistency does not arise.

It follows from Proposition 6 that if a discount function $\alpha$ satisfies $\alpha(0+) = 1$ and supp$\{\alpha\}$ is open in $R_+$ (e.g., these conditions hold if $\alpha$ is continuous), then all $RDPP^{(\alpha)}$-consistent SPOs coincide on the set of finite projects. Since real-world projects are finite, this means, roughly speaking, that there is a unique $RDPP^{(\alpha)}$-consistent SPO $\succeq$. This SPO has a straightforward interpretation: for finite projects $x$ and $y$, $x \succeq y \Leftrightarrow$ for every time $\tau \in R_+$, if project $y$ truncated at $\tau$ is profitable (i.e., has nonnegative NPV, $F^{(\alpha)}(y_{\leq \tau}) \geq 0$), then so is project $x$ truncated at $\tau$. In view of the interpretation, this SPO is a proper profitability measure if there is a possibility of early project termination for external reasons.

We proceed by considering a refinement of the adopted definition of DPP. We say that an SPO $\succeq$ is *stable under truncation* if $x \succeq y \Rightarrow x_{\leq \tau} \succeq y_{\leq \tau}$ for any $\tau \geq 0$. The condition states that truncation does not cause a perturbation of the profitability ordering; so SPOs satisfying this condition are particularly useful under possibility of early project termination. In order to describe SPOs that are stable under truncation we introduce the following notation. Given $\alpha \in \mathcal{A}$, denote by $G_{\tau,\lambda}^{(\alpha)}$, $\tau > 0$, $\lambda \in [0,1]$ the NPV functional induced by the discount function $\lambda \alpha I_{[0,\tau]} + (1-\lambda) \alpha I_{[0,\tau)}$. Note that $G_{\tau,\lambda}^{(\alpha)} = \lambda G_{\tau}^{(\alpha)} + (1-\lambda) \lim_{t \to \tau^-} G_t^{(\alpha)}$. The next proposition characterizes $Q_1'$-complete SPOs that are stable under truncation.

**Proposition 7.**

*Let $\succeq$ be an SPO with a representation $\mathcal{F}$. The following conditions are equivalent:*

(a) $\succeq$ *is $Q_1'$-complete and stable under truncation;*

(b) *there exists $\alpha \in \mathcal{A}$ such that*
$\{G_{\tau}^{(\alpha)}, \tau \in \{0\} \cup \mathrm{int}(\mathrm{supp}\{\alpha\})\} \subseteq \mathcal{F} \subseteq \{F^{(\chi)}, F^{(\alpha)}\} \cup \{G_{t,\lambda}^{(\alpha)}, (t,\lambda) \in (\mathrm{supp}\{\alpha\} \setminus \{0\}) \times [0,1]\}$.

Proposition 7 suggests the following refinement of the conventional definition of DPP. A project $x \in \mathrm{P}$ is said to *possess the refined DPP w.r.t. $\alpha$* if $x(0) < 0$ and there exists a number $\tau^{(\alpha)}(x) \in \mathrm{R}_{++}$ such that $G_{t,0}^{(\alpha)}(x) < 0$ and $G_t^{(\alpha)}(x) < 0$ for all $t \in (0, \tau^{(\alpha)}(x))$ and $G_t^{(\alpha)}(x) \geq 0$ for all $t \in [\tau^{(\alpha)}(x), +\infty)$. If $x$ possesses the refined DPP w.r.t. $\alpha$, then denote by $\lambda^{(\alpha)}(x)$ the least solution $\lambda \in [0,1]$ of the equation $G_{\tau^{(\alpha)}(x),\lambda}^{(\alpha)}(x) = 0$. Let $\succeq$ be the SPO induced by $\{F^{(\chi)}, F^{(\alpha)}\} \cup \{G_{t,\lambda}^{(\alpha)}, (t,\lambda) \in (\mathrm{supp}\{\alpha\} \setminus \{0\}) \times [0,1]\}$. Then, provided that projects $x$ and $y$ possess the refined DPP w.r.t. $\alpha$,

$$x \succeq y \Leftrightarrow (\tau^{(\alpha)}(x), \lambda^{(\alpha)}(x)) \leq_{\mathrm{lex}} (\tau^{(\alpha)}(y), \lambda^{(\alpha)}(y)), \tag{9}$$

where $\leq_{\mathrm{lex}}$ is the lexicographic order. Clearly, if $x$ possesses the refined DPP w.r.t. $\alpha$, it also possesses the DPP w.r.t. $\alpha$ and $DPP^{(\alpha)}(x) = \tau^{(\alpha)}(x)$. If $x$ is finite, the converse is also true.

Real-world investment projects are finite. It is a common practice to use linear interpolation of the cumulative discounted cash flow to evaluate DPP of a finite project (e.g., see Götze et al., 2015, p. 72). In particular, for a discrete project $x$, the DPP obtained via interpolation is given by $DPP_*^{(\alpha)}(x) := \tau^{(\alpha)}(x) - 1 + \lambda^{(\alpha)}(x)$. Note that the restriction of the ordering (9) to the set of discrete projects possessing the DPP w.r.t. $\alpha$ coincides with the ordering induced by $1/DPP_*^{(\alpha)}$. This observation provides a rigorous justification for the linear interpolation practice. Note that this justification has nothing to do with the assumption that the observed cash flow is the discretization of a continuous cash flow.

To illustrate the notion of refined DPP, consider projects $x = -2 \cdot 1_0 + 5 \cdot 1_1$ and $y = -1_0 + 2 \cdot 1_1$, where time is measured in years. Both have conventional PP 1 year. However, for the SPO $\succeq$ induced by $\{F^{(\chi)}, F^{(1_0)}\} \cup \{G_{t,\lambda}^{(1_0)}, (t,\lambda) \in \mathrm{R}_{++} \times [0,1]\}$, we have $x \succ y$ as $(\tau^{(1_0)}(x), \lambda^{(1_0)}(x)) = (1, 2/5)$ is lexicographically strictly smaller than $(\tau^{(1_0)}(y), \lambda^{(1_0)}(y)) = (1, 1/2)$ (or equivalently $DPP_*^{(1_0)}(x) = \tau^{(1_0)}(x) - 1 + \lambda^{(1_0)}(x) = 2/5 < 1/2 = \tau^{(1_0)}(y) - 1 + \lambda^{(1_0)}(y) = DPP_*^{(1_0)}(y)$).

### 4.3. PI and other ratio type indices

In this section, we show that the profitability index $PI^F$ introduced in Section 3 (as well as some other ratio type indices) is a profitability metric. We find its natural domain and the natural extension, describe corresponding consistent SPOs and provide their axiomatic characterization.

Let $F$ and $G$ ($\neq F$) be NPV functionals. A project $x \in P$ is said to *possess the ratio index* (*RI*) *w.r.t.* $F$ *and* $G$ if $F(x) \geq 0$ and $G(x) < 0$; in this case, the index is defined as $RI_G^F(x) := (F(x) - G(x))/(-G(x))$. $RI_G^F$ comprises several popular profitability measures. For instance, the undiscounted profitability index (the return on investment), $x \mapsto (x(+\infty) - x(0))/(-x(0))$, considered in Example 3, corresponds to $F = F^{(1_0)}$ and $G = F^{(\chi)}$. The discounted profitability index $PI^F(x) = (F(x) - x(0))/(-x(0))$ corresponds to $RI_G^F$ with $G = F^{(\chi)}$. If $x \in Q_2$, i.e., $x$ is a conventional investment in which a series of cash outflows is followed after some time $\tau \in R_+$ by a series of cash inflows, then $RI_G^F(x)$ is the discounted benefit-cost ratio provided that $F = F^{(\alpha)}$ and $G = G_\tau^{(\alpha)}$.

Denote by $Q_G^F \subset P$ the set of projects possessing the RI w.r.t. $F$ and $G$. Note that $Q_G^F \neq \emptyset$ for any distinct $F, G \in \mathcal{NPV}$. The next result shows that $RI_G^F : Q_G^F \to [1, +\infty)$ is a profitability metric, describes its natural domain, the natural extension, and $RI_G^F$-consistent SPOs.

**Proposition 8.**

*Given $F, G \in \mathcal{NPV}$, $F \neq G$, put $\tilde{W} := \{w \in R : wF + (1-w)G \in \mathcal{NPV}\}$, $\tilde{F} := G + (\sup \tilde{W})(F - G)$, $\tilde{G} := G + (\inf \tilde{W})(F - G)$.[16] The following statements hold.*

(a)  $RI_G^F : Q_G^F \to [1, +\infty)$ *is a profitability metric.*

(b)  *An SPO is $RI_G^F$-consistent if and only if it is induced by $\{wF + (1-w)G, w \in W\}$, $(0,1) \subseteq W \subseteq \tilde{W}$.*

(c)  *The least $RI_G^F$-consistent SPO is induced by $\{w\tilde{F} + (1-w)\tilde{G}, w \in [0,1]\}$.*

(d)  *The natural domain of $RI_G^F$ is $D_G^F = P \setminus \{x \in P : \tilde{F}(x) < 0, \tilde{G}(x) \geq 0\}$.*

(e)  *The natural extension of $RI_G^F$ is the complete preorder on $D_G^F$ with a utility representation $\overline{RI}_G^F : D_G^F \to \overline{R}_+$ given by*

$$\overline{RI}_G^F(x) := \begin{cases} 0 & \text{if } \tilde{F}(x) < 0 \text{ and } \tilde{G}(x) < 0 \\ RI_{\tilde{G}}^{\tilde{F}}(x) & \text{if } \tilde{F}(x) \geq 0 \text{ and } \tilde{G}(x) < 0 \\ +\infty & \text{if } \tilde{F}(x) \geq 0 \text{ and } \tilde{G}(x) \geq 0 \end{cases}.$$

As a corollary from Proposition 8, we derive the natural domain of the profitability index, its natural extension, and $PI^F$-consistent SPOs. Indeed, let $F$ be an NPV functional associated with a discount function $\alpha$ satisfying $\alpha(0+) = 1$. It follows from Proposition 8 with $G = F^{(\chi)}$ that the natural domain of $PI^F: \{x \in P: F(x) \geq 0, x(0) < 0\} \to [1, +\infty)$ is $P \setminus \{x \in P: F(x) < 0, x(0) \geq 0\}$, the natural extension is

$$\overline{PI}^F(x) := \begin{cases} 0 & \text{if } F(x) < 0 \text{ and } x(0) < 0 \\ PI^F(x) & \text{if } F(x) \geq 0 \text{ and } x(0) < 0 \\ +\infty & \text{if } F(x) \geq 0 \text{ and } x(0) \geq 0 \end{cases}$$

and $PI^F$-consistent SPOs are induced by $\{H_\gamma^{(\alpha)}, \gamma \in \Gamma\}$, $(0,1) \subseteq \Gamma \subseteq [0,1]$ (recall that $H_\gamma^{(\alpha)}$ is the NPV functional associated with the discount function $\gamma\alpha + (1-\gamma)\chi$).

---

[16] Since the set $\mathcal{NPV}$ is compact, so is $\tilde{W}$, and therefore, $\tilde{F}, \tilde{G} \in \mathcal{NPV}$.

For instance, let $F$ be the NPV functional associated with the discount function $1_0$, i.e., $PI^F$ is the undiscounted profitability index. Consider projects $x = 1_0 - 100 \cdot 1_1 + 120 \cdot 1_2$, $y = -25 \cdot 1_0 + 36 \cdot 1_2$, and $-y$. Project $x$ is adopted from Example 5; projects $y$ and $-y$ are the lender's and borrower's cash flows for a 2-year loan at 20% per year. As $PI^F(x)$ and $PI^F(-y)$ are undefined, projects $x$, $y$, and $-y$ are incomparable with respect to $PI^F$. Now let $\succeq$ be a $PI^F$-consistent SPO. We have $\{\gamma \in [0,1]: H_\gamma^{(1_0)}(x) \geq 0\} = [0,1]$, $\{\gamma \in [0,1]: H_\gamma^{(1_0)}(y) \geq 0\} = [25/36, 1]$, and $\{\gamma \in [0,1]: H_\gamma^{(1_0)}(-y) \geq 0\} = [0, 25/36]$. As $[0,1] \supset [25/36, 1]$ and $[0,1] \supset [0, 25/36]$, we get $x \succ y$, $x \succ -y$, whereas $y$ and $-y$ are incomparable.

### 4.4. Discussion

Based on the analysis above, we suggest to replace the conventional profitability metrics – IRR, PP/DPP, and PI – with their extensions induced by the corresponding (least) consistent SPOs described, respectively, in Propositions 3, 6, and 8. These extensions enjoy the same natural properties as the underlying profitability metrics but substantially enlarge the set of comparable projects.

In the capital budgeting literature, the NPV criterion is unanimously accepted as the gold standard, whereas other profitability metrics are often put in a subordinated position, though can complement the criterion with important pieces of information. Our analysis shows that each profitability metric plays its own role in the presence of uncertainty. The interpretation of an SPO (as a measure of a project's financial stability across various economic scenarios) suggests that the choice of a particular metric should be determined by the source of uncertainty an investor faces and hence provides a clear condition under which the application of one metric is superior to others.

For instance, assume that the investor faces the possibility of early project termination for external environmental reasons and the project can be terminated at any date $\tau \in R_{++}$. Given a discount function $\alpha$, this results in the set of scenarios represented by the collection $\mathcal{F} = \{G_\tau^{(\alpha)}, \tau \in R_{++}\} \cup \{F^{(\alpha)}\}$ of NPV functionals (recall that $G_\tau^{(\alpha)}(x)$ is the NPV of project $x$ truncated at $\tau$). The SPO induced by that $\mathcal{F}$ is $RDPP^{(\alpha)}$-consistent (Proposition 6), so DPP with the discount function $\alpha$ is an appropriate tool to evaluate project profitability in this particular case. As another example, assume that the investor faces complete uncertainty with respect to the cost of capital and any nonnegative cost of capital is plausible. This results in the set $\mathcal{F}^{(E)}$ of NPV functionals. As shown in Proposition 3, the SPO induced by $\mathcal{F}^{(E)}$ is $IRR^{(E)}$-consistent, so the conventional IRR is an appropriate tool to evaluate project profitability in this example. In the same manner, analyzing the structure of a $PI^F$-consistent SPO, we conclude that the profitability index should be used under uncertainty with respect to the future cash flow, $x - x(0)1_0$, in the form of its proportional (to some unknown scale factor $\gamma \in (0,1]$) reduction. Indeed, given a discount function $\alpha \neq \chi$, this source of uncertainty results in the collection $\mathcal{F} = \{x \mapsto F^{(\alpha)}(x(0)1_0 + \gamma(x - x(0)1_0)), \gamma \in (0,1]\} = \{H_\gamma^{(\alpha)}, \gamma \in (0,1]\}$ of NPV functionals. The SPO induced by that $\mathcal{F}$ is $PI^F$-consistent with $F = F^{(\alpha)}$ (Proposition 8), so the profitability index $PI^F$ is an appropriate tool to evaluate project profitability in this particular case. Finally, the structure of the set representing the NPV criterion – a singleton – shows that it should be used under complete certainty. While some of these recommendations are already used in practice, our analysis provides a formal justification for them.

Since there are other sources of uncertainty, the collection of conventional metrics – IRR, PP/DPP, and PI – cannot be considered as comprehensive. For instance, investment in the real sector may suffer from uncertain intensity of project implementation. Given $\alpha \in \mathcal{A}$, this results in the collection $\{U_\lambda^{(\alpha)}, \lambda \in R_{++}\}$ of valuation functionals, where $U_\lambda^{(\alpha)}(x) := x(0) + \int_0^\infty \alpha(t) dx(\lambda t)$ is the value of project $x$ implemented with the intensity $\lambda$. Changing the variable in the integral, we get that $U_\lambda^{(\alpha)}$ is the NPV functional associated with the discount function $t \mapsto \alpha(t/\lambda)$. Therefore, the SPO induced by $\{U_\lambda^{(\alpha)}, \lambda \in R_{++}\}$ can be used to evaluate profitability if the investor faces uncertain intensity of project implementation. Note that if $\alpha(t) = e^{-\lambda_0 t}$, $\lambda_0 \in R_{++}$, then this SPO is $IRR^{(E)}$-consistent, so the resulting profitability metric reduces to the conventional IRR.

Investment in the real sector may also face the possibility of postponement of the project implementation, say, due to a pandemic. Given $\alpha \in \mathcal{A}$, this results in the collection $\{V_\tau^{(\alpha)}, \tau \in R_+\}$ of valuation functionals, where $V_\tau^{(\alpha)}(x) := x(0) + \int_\tau^\infty \alpha(t) dx(t - \tau)$, $\tau \in R_+$ is the value of project $x$ whose implementation (with the exception of the initial transaction, $x(0)$) is postponed until $\tau$. Changing the variable in the integral, we get that $V_\tau^{(\alpha)}$ is the NPV functional associated with the discount function $t \mapsto \chi(t) + \alpha(t + \tau)I_{(0,+\infty)}(t)$. Therefore, the SPO induced by $\{V_\tau^{(\alpha)}, \tau \in R_+\}$ can be used if the investor faces the possibility of project postponement. Note that if $\alpha(t) = e^{-\lambda t}$, $\lambda \in R_{++}$, then $\{V_\tau^{(\alpha)}, \tau \in R_+\} = \{H_\gamma^{(\alpha)}, \gamma \in (0,1]\}$, so the resulting profitability metric is the conventional PI.

Some authors argue to use multiple criteria (say, the combination of IRR and DPP) to choose between projects. In view of property 6 in Lemma 1, such a multiple criterion can be constructed by uniting the sets of NPV functionals representing the partial criteria.

To summarize, we suggest to replace the zoo of conventional capital budgeting techniques with context-specific SPOs determined by sources of uncertainty the investor faces. For instance, assume that the investor faces an uncertain cost of capital in the range of 2–4% and the possibility of project termination in 5 years or later due to climate change. Then, the SPO with a representation induced by the set of discount functions $\{t \mapsto (1+\lambda)^{-t} I_{[0,\tau]}(t), \lambda \in [0.02, 0.04], \tau \in [5,+\infty)\}$ is a proper tool to evaluate project profitability in this case. As another example, as is known, the conventional IRR is incompatible with time-varying cost of capital. To overcome this problem, we suggest to use the SPO induced by a set of discount functions representing plausible term structure patterns for the cost of capital (note that this SPO produces the same ordering as IRR in the case of constant unknown cost of capital discussed above). For instance, suppose that the cost of capital, which is currently 20%, is expected to grow at a time-varying rate that does not exceed 1% per year. Then, the SPO with a representation induced by the set of discount functions $\{t \mapsto \exp\left(-\int_0^t \lambda(\tau)\mathrm{d}\tau\right) : \lambda(0) = 0.2,\ 0 \leq \lambda'(\tau) \leq 0.01\ \forall \tau \in \mathrm{R}_+\}$ is an appropriate tool to evaluate project profitability in this case.

Several conventional performance measures do not satisfy our axioms of a profitability metric. For instance, so do all known modifications and extensions (but not the restrictions) of the conventional IRR, the discounted benefit-cost ratio (the ratio of the total discounted cash benefits to the total discounted cash costs), and PP/DPP defined in Hajdasiński (1993). The investor must be careful when using them to set a target level of profitability in the presence of several independent investment proposals: implementation of those proposals that meet the target does not guarantee the resulting pool of proposals to achieve the target.

## 5. Conclusion

This paper provides an axiomatic foundation for a profitability ranking of investment projects. We adopt axioms similar to those used in Promislow (1997) and Vilensky and Smolyak (1999), but in contrast to the latter paper, allow incomparable projects. This results in a class of orderings that includes those induced by conventional capital budgeting metrics, in particular, IRR, PP, DPP, and PI.

The project space $\mathrm{P}$ we deal with covers investment projects with bounded deterministic cash flows. Theoretical financial models operate unbounded and/or stochastic cash flows, so other types of project spaces are of interest. Note that all the obtained results that do not explicitly rely on the structure of an NPV functional (namely, Propositions 1, 2, 8 and Lemmas 1, 2, 6, 7, 9) remain valid for an ordered Hausdorff locally convex topological vector space with an order unit. That is, if $\mathrm{P}$ is a real Hausdorff locally convex topological vector space, $\mathrm{P}_+ \subset \mathrm{P}$ is a closed convex cone with a nonempty interior $\mathrm{P}_{++} = \mathrm{int}\,\mathrm{P}_+$, and $\mathcal{NPV} = \{F \in \mathrm{P}_+^\circ : F(e) = 1\}$, where $e \in \mathrm{P}_{++}$ is a distinguished element called an order unit. Notice that the existence of an order unit implies that the cone of nonnegative projects $\mathrm{P}_+$ is generating, i.e., every $x \in \mathrm{P}$ can be represented in the form $x = x_+ - x_-$, $x_+, x_- \in \mathrm{P}_+$. Such a representation is vital for $x$ to be interpreted as a cash flow as, by definition, cash flow is the net of cash inflows and outflows.

We close with a discussion of open problems and directions for future research.
1. The paper mainly exploits SPO (rather than PO) due to its simple representation and nice interpretation. Therefore, it would be desirable to provide its separate axiomatic characterization.
2. A slightly more intuitive relation than an SPO can be introduced as follows. Given a nonempty set $\mathcal{F} \subseteq \mathcal{NPV}$, define the preorder $\succsim$ on $\mathrm{P}$ by
$$x \succsim y \iff \mathrm{sgn}\,F(x) \geq \mathrm{sgn}\,F(y) \text{ for all } F \in \mathcal{F}.$$
The relation $\succsim$ seems to be a little bit more relevant for profitability measurement purposes than the SPO $\succeq$ induced by $\mathcal{F}$. First, $x \succ 0 \succ -x$ (where $\succ$ is the asymmetric part of $\succsim$) for every $x \in \mathrm{P}_{++}$, whereas for $\succeq$, we have a counterintuitive $0 \succeq x$ for all $x \in \mathrm{P}$. Second, in contrast to $\succeq$, $\succsim$ satisfies the skew symmetry condition, $x \succsim y \implies -y \succsim -x$. Various types of projects imply the existence of two sides, whose cash flows differ by sign (e.g., the borrower and lender sides of a loan). The skew symmetry condition asserts that the two sides rank projects' profitabilities in the reversed order. Although the preorders $\gtrsim$ and $\succeq$ are "essentially the same" from a practical viewpoint (the closure of an upper contour set of $\gtrsim$ coincides with the corresponding upper contour set of $\succeq$), a natural progression of this work is to analyze $\gtrsim$ and exploit it to study profitability metrics in the manner of Sections 3 and 4.

3. Many capital budgeting techniques use various primitives besides or instead of cash flow stream. For instance, the average IRR paradigm (Magni, 2010, 2016) is based on income and capital stream. The notions of PO, SPO, and profitability metric can, with obvious changes, be defined on the set of pairs ⟨income, capital stream⟩, rather than on the set of cash flow streams. It is an interesting direction for future research to study these objects in the manner of Sections 2–4.

## 6. Appendix: auxiliary results and proofs

**Lemma 8.**

$F : \mathrm{P} \to \mathrm{R}$ *is a net present value functional, i.e.,* $F \in \mathcal{NPV}$, *if and only if representation (1) holds for some* $\alpha \in \mathcal{A}$.

**Proof.**

Let $F \in \mathcal{NPV}$. Since $\mathrm{int}\, \mathrm{P}_+ = \mathrm{P}_{++} \neq \varnothing$, an additive and positive functional on $\mathrm{P}$ is homogeneous and continuous (Jameson, 1970, Corollary 3.1.4, p. 81). Therefore, there exists a function of bounded variation $\alpha : \mathrm{R}_+ \to \mathrm{R}$ such that $F(x) = \alpha(0)x(0) + \int_0^\infty \alpha(t)\mathrm{d}x(t)$ (Monteiro et al., 2018, Theorem 8.2.8, p. 304). The function $\alpha$ is nonnegative: indeed, for any $t \in \mathrm{R}_+$, we have $1_t \in \mathrm{P}_+$ and therefore $\alpha(t) = F(1_t) \geq 0$. $\alpha$ is nonincreasing: for any $t < \tau$, we have $\alpha(t) - \alpha(\tau) = F(1_t) - F(1_\tau) = F(1_t - 1_\tau) \geq 0$ as $1_t - 1_\tau \in \mathrm{P}_+$. Clearly, $\alpha(0) = F(1_0) = 1$, so $\alpha \in \mathcal{A}$.

Now assume that (1) holds for some $\alpha \in \mathcal{A}$. Clearly, $F \in \mathrm{P}^*$ and $F(1_0) = 1$, so we only have to prove that $F$ is positive. Pick $x \in \mathrm{P}_+$ and note that for any $\varepsilon > 0$, there is a step-function $y = \sum_{k=1}^n c_k 1_{t_k} \in \mathrm{P}$, $c_1, ..., c_n \in \mathrm{R}$, $0 \leq t_1 < ... < t_n$ such that $\|x - y\| < \varepsilon$ (Monteiro et al., 2018, p. 82). The constants $c_1, ..., c_n$ can be chosen such that $y \in \mathrm{P}_+$, i.e., $c_1 + ... + c_k \geq 0$ for all $k = 1, ..., n$; indeed, the step-function $y_+(t) := \max\{y(t), 0\}$ satisfies $y_+ \in \mathrm{P}_+$ and $\|x - y_+\| < \varepsilon$. As $\alpha \in \mathcal{A}$, we have

$$F(y) = \sum_{k=1}^n c_k \alpha(t_k) = \alpha(t_n)(c_1 + ... + c_n) + \sum_{k=1}^{n-1}(\alpha(t_k) - \alpha(t_{k+1}))(c_1 + ... + c_k) \geq 0.$$

Since $F$ is continuous, this proves that $F(x) \geq 0$. ∎

**Proof of Lemma 1.**

1. NP, INT, and USC imply $\lambda x \succeq x$, $\lambda > 0$. This holds for all $x \in \mathrm{P}$ and $\lambda > 0$ if and only if $\lambda x \sim x$.

2. Assume by way of contradiction that $\succeq$ is lower semicontinuous. Pick $x \in P$. By property 1, $x \sim \lambda x$ for all $\lambda > 0$. Tending $\lambda \to 0$ and using upper and lower semicontinuity of $\succeq$, we get $x \sim 0$, which contradicts nontriviality of $\succeq$.

3. By property 1, $2 \cdot 1_0 \sim 1_0$, whereas $2 \cdot 1_0 > 1_0$.

4. Since $1_0$ is an order unit and $\succeq$ is nontrivial, property 1 and MON imply $2 \cdot 1_0 \succ -1_0$. Now assume by way of contradiction that for all $x, y \in P$, $x \succ y \Rightarrow x \succ x + y$. Applying this implication to the inequality $2 \cdot 1_0 \succ -1_0$, we arrive at a contradiction with property 1: $2 \cdot 1_0 \succ 2 \cdot 1_0 - 1_0 = 1_0$. The remaining statement can be established in a similar fashion.

5. Assume by way of contradiction that there is $x \in P$ such that $1_0 \succ x$, $1_0 \succ -x$, and $x \succeq -x$. As $L_\succeq(x)$ is closed under addition and $x, -x \in L_\succeq(x)$, we arrive at a contradiction: $x \succeq x - x = 0 \sim 1_0$.

6. Straightforward. ∎

**Proof of Proposition 1.**

To show the independence of NP, MON, INT, and USC, we provide four examples of binary relations on P that satisfy three of the conditions while violating the fourth. Pick $F \in \mathcal{NPV}$ and $G \in P^* \setminus P_+^\circ$. The binary relation $\succeq$ defined by $x \succeq y \Leftrightarrow \max\{I_{\{F\}^\circ}(x), I_{\{F\}^\circ}(y)\} = 1$ satisfies all the conditions except NP. The binary relation given by $x \succeq y \Leftrightarrow I_{\{G\}^\circ}(x) \geq I_{\{G\}^\circ}(y)$ meets all the conditions except MON. The binary relation given by $x \succeq y \Leftrightarrow F(x) \geq F(y)$ satisfies all the conditions except INT. Finally, the binary relation defined by $x \succeq y \Leftrightarrow I_{\{F\}^\circ}(-y) \geq I_{\{F\}^\circ}(-x)$ meets all the conditions except USC.

(a)$\Rightarrow$(b). Let $\succeq$ be a PO. From USC, INT, and property 1 in Lemma 1, it follows that for any $z \in P$, $U_\succeq(z)$ is a closed convex cone. Set $\mathcal{U} = \{(U_\succeq(z))^\circ \cap \mathcal{NPV}, z \in P\}$. By MON, $(U_\succeq(z))^\circ \cap \mathcal{NPV}$ is a base for the cone $(U_\succeq(z))^\circ$. Thus, $((U_\succeq(z))^\circ \cap \mathcal{NPV})^\circ = (U_\succeq(z))^{\circ\circ} = U_\succeq(z)$, where the second equality follows from the bipolar theorem (Aliprantis and Border, 2006, Theorem 5.103, p. 217). By NP, $x \succeq y \Leftrightarrow \{z \in P : x \in U_\succeq(z)\} \supseteq \{z \in P : y \in U_\succeq(z)\} \Leftrightarrow \{K \in \mathcal{U} : x \in K^\circ\} \supseteq \{K \in \mathcal{U} : y \in K^\circ\} \Leftrightarrow I_{K^\circ}(x) \geq I_{K^\circ}(y)$ for all $K \in \mathcal{U}$. From INT it follows that the set $L_\succeq(z) = \{x \in P : z \succeq x\} = \bigcap_{K \in \mathcal{U}: z \notin K^\circ} (P \setminus K^\circ)$ is closed under addition.

(b)$\Rightarrow$(a). It is straightforward to verify that the binary relation $\succeq$ defined in part (b) is a PO.

In order to show that elements of the family $\mathcal{U}$ in part (b) can be chosen closed and convex, note that $\succeq$ depends on $K \in \mathcal{U}$ only through $K^\circ$. For each $K \in \mathcal{U}$, set $\bar{K} := K^{\circ\circ} \cap \mathcal{NPV}$. We have $\bar{K}^\circ = (K^{\circ\circ})^\circ = (K^\circ)^{\circ\circ} = K^\circ$, where the first equality follows from the fact that $\bar{K}$ is a base for the cone $K^{\circ\circ}$ and the last equality comes from the bipolar theorem. Thus, replacing each $K \in \mathcal{U}$ with $\bar{K}$ in the representation produces the same PO. $\bar{K}$ is closed and convex as the intersection of the closed and convex sets $K^{\circ\circ}$, $P_+^\circ$, and $\{F \in P^* : F(1_0) = 1\}$. ∎

**Proof of Proposition 2.**

(a)$\Rightarrow$(b), (a)$\Rightarrow$(c), (a)$\Rightarrow$(d). Straightforward.

(b)$\Rightarrow$(a). Set $U := \{z \in P : z \sim 1_0\}$. Property 1 in Lemma 1, MON, and NP imply that $U = U_{\succeq}(1_0)$, so, by INT and USC, $U$ is a closed convex cone. Set $L := P \setminus U$. Since $\succeq$ is nontrivial, $L \neq \varnothing$. Pick $x, y \in L$. As $\succeq$ is complete, without loss of generality we may assume that $x \succeq y$. Combining this with $x \succeq x$ and using INT, we get $x \succeq x + y$ and therefore $x + y \in L$. This proves that $L$ is an open convex cone. By a separating hyperplane theorem (Aliprantis and Border, 2006, Lemma 5.66, Theorem 5.67, p. 202), there is a nonzero $F \in P^*$ such that $F(x) < 0 \leq F(y)$ for all $x \in L$ and $y \in U$. Condition MON implies $P_+ \subseteq U$, so $F$ can be chosen such that $F \in \mathcal{NPV}$. Since $P = L \cup U$, we have $U = \{F\}^\circ$. For each $z \in L$, $U_{\succeq}(z)$ is a closed convex cone containing $U$ as a proper subset. As $U$ is a closed half-space, $U_{\succeq}(z) = P$. Thus, $L$ is an equivalence class w.r.t. $\sim$, so $x \succeq y \Leftrightarrow I_{\{F\}^\circ}(x) \geq I_{\{F\}^\circ}(y)$.

(c)$\Rightarrow$(a). Reproducing the beginning of the proof "(b)$\Rightarrow$(a)", we get that $U := \{z \in P : z \sim 1_0\} = U_{\succeq}(1_0)$ is a closed convex cone and $L := P \setminus U \neq \varnothing$. Condition (c) implies that $L$ is an open convex cone. The rest of the proof reproduces the corresponding part of that of "(b)$\Rightarrow$(a)".

(d)$\Rightarrow$(a). Set $L := \{z \in P : z \sim -1_0\}$ and $U := U_{\succ}(-1_0)$. From property 1 in Lemma 1, MON, and NP it follows that $U = P \setminus L$. Nontriviality of $\succeq$, property 1, and INT (resp. condition (d)) imply that $L$ (resp. $U$) is a nonempty convex cone. By a separating hyperplane theorem, there is a nonzero $F \in P^*$ such that $F(x) \leq 0 \leq F(y)$ for all $x \in L$ and $y \in U$. As $P_+ \subseteq U$, $F$ can be chosen such that $F \in \mathcal{NPV}$. Set $H_+ := \{x \in P : F(x) > 0\}$ and $H_- := \{x \in P : F(x) < 0\}$ and note that $H_+ \subseteq U$ and $H_- \subseteq L$. For each $x \in H_+$, $L_{\succeq}(x)$ is a convex cone containing $L \cup \{x\}$, i.e., $R_+ x + L \subseteq L_{\succeq}(x)$. As $R_+ x + L \supseteq R_+ x + H_- = P$, we get $L_{\succeq}(x) = P$. This proves that $H_+ \subseteq U_{\succeq}(1_0)$. We have $\{F\}^\circ = \mathrm{cl}(H_+) \subseteq U_{\succeq}(1_0) \subseteq U \subseteq \{F\}^\circ$, where the first inclusion follows from the fact that $U_{\succeq}(1_0)$ is closed. Thus, $U = U_{\succeq}(1_0) = \{F\}^\circ$. As $U_{\succeq}(1_0) = \{z \in P : z \sim 1_0\}$, i.e., $U$ is an equivalence class w.r.t. $\sim$, we are done. ∎

**Lemma 9.**

*Given a nonempty set $\mathcal{S} \subseteq \mathcal{NPV}$, let $\mathcal{F}$ be the closed (in the weak* topology) convex hull of $\mathcal{S}$ and $\succeq$ be the SPO induced by $\mathcal{F}$. The following conditions are equivalent:*

(a) $x \succeq y$;

(b) $\sup_{\lambda \in R_+} \inf_{F \in \mathcal{S}} F(x - \lambda y) \geq 0$;

(c) $x \in \mathrm{cl}(\mathcal{S}^\circ + (R_+ y))$.

*In particular, if the set $\mathcal{S}^\circ + (\mathrm{R}_+ y)$ is closed (which holds, e.g., if $\mathcal{S}$ is finite[17]), then (a)–(c) are also equivalent to*

(d) *there exists $\lambda \in \mathrm{R}_+$ such that $F(x - \lambda y) \geq 0$ for all $F \in \mathcal{S}$.*

**Proof.**

Note that condition (b) is equivalent to the following one which we refer to as (b)': for any $\varepsilon > 0$, there exists $\lambda \in \mathrm{R}_+$ such that $F(x - \lambda y) + \varepsilon \geq 0$ for all $F \in \mathcal{S}$. Condition (c) is equivalent to the following one which we refer to as (c)': for any neighborhood of zero $O$ in $\mathrm{P}$, there exists $(\lambda, z) \in \mathrm{R}_+ \times O$ such that $F(x + z - \lambda y) \geq 0$ for all $F \in \mathcal{S}$.

(b)'$\Rightarrow$(c)'. Pick an open neighborhood of zero $O$ in $\mathrm{P}$. Since $O$ is absorbing, we have $\varepsilon 1_0 \in O$ for some $\varepsilon > 0$. By (b)', there exists $\lambda \in \mathrm{R}_+$ such that $F(x - \lambda y) + \varepsilon \geq 0$ for all $F \in \mathcal{S}$. Thus, (c)' holds with that $\lambda$ and $z = \varepsilon 1_0$.

(c)'$\Rightarrow$(b)'. Pick $\varepsilon > 0$ and set $O_\varepsilon := \varepsilon 1_0 - \mathrm{P}_{++}$. As $O_\varepsilon$ is an open neighborhood of zero, condition (c)' implies that there exists $(\lambda, z) \in \mathrm{R}_+ \times O_\varepsilon$ such that $F(x + z - \lambda y) \geq 0$ for all $F \in \mathcal{S}$. Note that $s \in \mathrm{P}_{++}$ if and only if $F(s) > 0$ for all $F \in \mathcal{NPV}$ (Aliprantis and Tourky, 2007, Lemma 2.17, p. 73). As $\varepsilon 1_0 - z \in \mathrm{P}_{++}$, we have $\varepsilon = F(\varepsilon 1_0) > F(z)$ for all $F \in \mathcal{NPV}$. Thus, $F(x - \lambda y) + \varepsilon > F(x + z - \lambda y) \geq 0$ for all $F \in \mathcal{S}$.

(a)$\Leftrightarrow$(c). Since $\mathrm{int}\,\mathrm{P}_+ = \mathrm{P}_{++} \neq \varnothing$, the set $\mathcal{NPV}$ is compact (Jameson, 1970, Theorem 3.8.6, p. 123) and therefore so is $\mathcal{F}$. $\mathcal{F}$ constitutes a compact base for the cone $\mathrm{R}_+ \mathcal{F}$ generated by $\mathcal{F}$, so $\mathrm{R}_+ \mathcal{F}$ is closed (Jameson, 1970, Theorem 3.8.3, p. 121). Thus, $\mathrm{R}_+ \mathcal{F}$ is the closed convex conical hull of $\mathcal{S}$ (i.e., the smallest closed convex cone containing $\mathcal{S}$) and $\mathcal{S}^{\circ\circ} = \mathrm{R}_+ \mathcal{F}$ by the bipolar theorem. We have

$$\mathrm{U}_{\succeq}(y) = (\mathcal{F} \cap \{y\}^\circ)^\circ = ((\mathrm{R}_+\mathcal{F}) \cap (\mathrm{R}_+ y)^\circ)^\circ = (\mathcal{S}^{\circ\circ} \cap (\mathrm{R}_+ y)^\circ)^\circ = (\mathcal{S}^\circ + (\mathrm{R}_+ y))^{\circ\circ} = \mathrm{cl}(\mathcal{S}^\circ + (\mathrm{R}_+ y)),$$

where the first equality comes from the definition of the SPO induced by $\mathcal{F}$, whereas the remaining equalities follow from the properties of the duality operation (e.g., see Messerschmidt, 2015, Lemma 2.1) and the fact that the initial and the weak topologies on $\mathrm{P}$ have the same collection of closed convex sets (Aliprantis and Border, 2006, Theorem 5.98, p. 214).

(c)$\Leftrightarrow$(d). Trivial. ∎

**Proof of Lemma 2.**

Assume that $\succeq$ is Q-complete. To show that $\succcurlyeq$ is complete, pick $F, G \in \mathcal{F}$ and assume by way of contradiction that neither $F \succcurlyeq G$ nor $G \succcurlyeq F$, i.e., there are $x, y \in \mathrm{Q}$ such that $x \notin \{F\}^\circ$, $x \in \{G\}^\circ$, $y \in \{F\}^\circ$, and $y \notin \{G\}^\circ$. This implies that $x$ and $y$ are incomparable w.r.t. $\succeq$, which is a contradiction. The same argument works in the other direction. ∎

**Proof of Lemma 3.**

---

[17] If $\mathcal{S}$ finite, then for any $y \in \mathrm{P}$, the cone $\mathcal{S}^\circ + (\mathrm{R}_+ y)$ is polyhedral (Luan and Yen, 2020, Theorem 2.11) and therefore closed.

(a)⇒(b). Trivial.

(b)⇒(c). Let $\succcurlyeq$ be the preorder on $\mathcal{F}$ induced by $Q_1'$. By condition (b) and Lemma 2, $\succcurlyeq$ is complete, so it is sufficient to verify that for any $F, G \in \mathcal{F}$, $F \succcurlyeq G \Rightarrow F \succcurlyeq_1 G$. Pick $F, G \in \mathcal{F}$ and denote by $\alpha$ and $\beta$ the discount functions associated with $F$ and $G$. Without loss of generality, we may assume that $F \succcurlyeq G$. If $\beta = \chi$, then trivially $\alpha \geq \beta$. Otherwise, pick $t \in \text{supp}\{\beta\} \setminus \{0\}$ and set $x = -1_0 + (1/\beta(t))1_t$. Then $x \in Q_1'$ and $G(x) = 0$, so, by the definition of $\succcurlyeq$, we have $0 \leq F(x) = -1 + (1/\beta(t))\alpha(t)$ as desired.

(c)⇒(a). Let $\succcurlyeq$ be the preorder on $\mathcal{F}$ induced by $Q_1$. In view of Lemma 2, it is sufficient to show that for any $F, G \in \mathcal{F}$, $F \succcurlyeq_1 G \Rightarrow F \succcurlyeq G$. Pick $F, G \in \mathcal{F}$ and denote by $\alpha$ and $\beta$ the discount functions associated with $F$ and $G$. Assume that $F \succcurlyeq_1 G$ and pick $x \in Q_1$ such that $G(x) \geq 0$. We have to show that $F(x) \geq 0$. Since $x$ is nondecreasing and $\alpha \geq \beta$, we have

$$F(x) = x(0) + \int_0^\infty \alpha(t)\mathrm{d}x(t) \geq x(0) + \int_0^\infty \beta(t)\mathrm{d}x(t) = G(x) \geq 0. \blacksquare$$

**Proof of Lemma 4.**

(a)⇒(b). Trivial.

(b)⇒(c). Let $\succcurlyeq$ be the preorder on $\mathcal{F}$ induced by $Q_2'$. By condition (b) and Lemma 2, $\succcurlyeq$ is complete, so it is sufficient to verify that for any $F, G \in \mathcal{F}$, $F \succcurlyeq G \Rightarrow F \succcurlyeq_2 G$. Pick $F, G \in \mathcal{F}$ and denote by $\alpha$ and $\beta$ the discount functions associated with $F$ and $G$.

First, we show that $\text{supp}\{\alpha\} = \text{supp}\{\beta\}$. Pick $\tau > 0$ and assume by way of contradiction that $\alpha(\tau) = 0$, whereas $\beta(\tau) > 0$. Put $x = -1_0 + (1/\beta(\tau))1_\tau$ and $y = -1_\tau + c1_t$, where $\tau < t$ and $c \in (0, \beta(\tau)/\beta(t))$ (with the convention $\beta(\tau)/0 = +\infty$). Then $x, y \in Q_2'$, $F(x) < 0$, $F(y) = 0$, whereas $G(x) = 0$, $G(y) < 0$, so $x$ and $y$ are incomparable, which is a contradiction. This proves that $\text{supp}\{\beta\} \subseteq \text{supp}\{\alpha\}$. The reverse inclusion can be shown in a similar fashion.

If $\text{supp}\{\beta\} = \{0\}$, then $\mathcal{F}$ is a singleton, and (c) trivially holds. Thus, it remains to consider the case $\text{supp}\{\beta\} \neq \{0\}$. Without loss of generality, we may assume that $F \succcurlyeq G$. Pick $0 \leq t < \tau \in \text{supp}\{\beta\}$ and set $x = -1_t + (\beta(t)/\beta(\tau))1_\tau$. Then $x \in Q_2'$ and $G(x) = 0$, so, by the definition of $\succcurlyeq$, we have $0 \leq F(x) = -\alpha(t) + (\beta(t)/\beta(\tau))\alpha(\tau)$ as desired.

(c)⇒(a). Let $\succcurlyeq$ be the preorder on $\mathcal{F}$ induced by $Q_2$. In view of Lemma 2, it is sufficient to verify that for any $F, G \in \mathcal{F}$, $F \succcurlyeq_2 G \Rightarrow F \succcurlyeq G$. Pick $F, G \in \mathcal{F}$ and let $\alpha$ and $\beta$ be the discount functions associated with $F$ and $G$. Assume that $F \succcurlyeq_2 G$ and pick $x \in Q_2$ such that $G(x) \geq 0$. We have to show that $F(x) \geq 0$. If $x \in Q_1$, the result follows from the fact that $\succcurlyeq_2 \subset \succcurlyeq_1$ and Lemma 3. Now assume that $x \in Q_2 \setminus Q_1$, i.e., $x(0) \leq 0$ and there is $\tau \in \mathbb{R}_{++}$ such that $x$ is nonincreasing (resp. nondecreasing) on $[0, \tau)$ (resp. $[\tau, +\infty)$). If $\tau \notin \text{supp}\{\beta\}$, the inequality $G(x) \geq 0$ implies that the restriction of $x$ to $\text{supp}\{\beta\}$ is identically 0 and, as $\text{supp}\{\alpha\} = \text{supp}\{\beta\}$, we have $F(x) = G(x) = 0$. Now assume that $\tau \in \text{supp}\{\beta\}$ and set $\Delta x(\tau) := x(\tau) - x(\tau-)$, $\tilde{x} := x - \Delta x(\tau)1_\tau$. We have

$$F(x) = x(0) + \int_0^\infty \alpha(t)\mathrm{d}x(t) = x(0) + \int_0^\infty \alpha \mathrm{d}(\tilde{x}(t) + \Delta x(\tau)1_\tau(t))$$

$$= x(0) + \int_0^\tau \alpha(t)d\tilde{x}(t) + \int_\tau^\infty \alpha(t)d\tilde{x}(t) + \alpha(\tau)\Delta x(\tau)$$

$$= x(0) + \frac{\alpha(\tau)}{\beta(\tau)}\left(\int_0^\tau \frac{\beta(\tau)}{\alpha(\tau)}\alpha(t)d\tilde{x}(t) + \int_\tau^\infty \frac{\beta(\tau)}{\alpha(\tau)}\alpha(t)d\tilde{x}(t) + \beta(\tau)\Delta x(\tau)\right)$$

$$\geq x(0) + \frac{\alpha(\tau)}{\beta(\tau)}\left(\int_0^\tau \beta(t)d\tilde{x}(t) + \int_\tau^\infty \beta(t)d\tilde{x}(t) + \beta(\tau)\Delta x(\tau)\right) = x(0) + \frac{\alpha(\tau)}{\beta(\tau)}(G(x) - x(0)) \geq G(x) \geq 0.$$

Here the first inequality stems from the facts that $t \mapsto \alpha(t)/\beta(t)$ is nondecreasing on supp$\{\beta\}$ (as $F \succcurlyeq_2 G$), $\tilde{x}$ is nonincreasing on $[0,\tau]$ and nondecreasing on $[\tau,+\infty)$. The second inequality follows from $G(x) - x(0) \geq 0$ (as $G(x) \geq 0$ and $x(0) \leq 0$).

(b)$\Rightarrow$(d). Trivial.

(d)$\Rightarrow$(c). The proof of "(b)$\Rightarrow$(c)" remains valid with $Q_2'$ replaced by $Q_2''$, provided that each NPV functional from $\mathcal{F}$ has a positive discount function. ∎

**Proof of Lemma 5.**

(a)$\Rightarrow$(b). Trivial.

(b)$\Rightarrow$(c). Let $\succeq$ be $Q_3'$-complete. Set $\alpha(t) := \sup_{F \in \mathcal{F}} F(1_t)$. Clearly, $\alpha \in \mathcal{A}$. Pick $F, G \in \mathcal{F}$ and denote by $\beta$ and $\delta$ the discount functions associated with $F$ and $G$. Let $\succcurlyeq$ be the preorder on $\mathcal{F}$ induced by $Q_3'$. By Lemma 2, $\succcurlyeq$ is complete. Without loss of generality, we may assume that $F \succcurlyeq G$. As $Q_1' \subset Q_3'$, $\succeq$ is $Q_1'$-complete, and, therefore, $\beta \geq \delta$ pointwise (Lemma 3). Pick $0 < t < \tau$ and set $x = -1_0 + a1_t + b1_\tau$. By the definition of $\succcurlyeq$, we must have $F(x) \geq 0$ for every $a$ and $b$ satisfying $G(x) = 0$. This condition implies $\det\begin{pmatrix} \beta(t) & \beta(\tau) \\ \delta(t) & \delta(\tau) \end{pmatrix} = 0$. Therefore, there is a constant $\gamma \in [0,1]$ such that $\delta = \gamma\beta + (1-\gamma)\chi$. If $\succeq$ is $Q_4'$-complete, then, as $Q_2' \subset Q_4'$, supp$\{\beta\}$ = supp$\{\delta\}$ (Lemma 4), so $\gamma \in (0,1]$. Therefore, $\mathcal{F} = \{H_\gamma^{(\alpha)}, \gamma \in \Gamma\}$ for some $\Gamma \subseteq [0,1]$ (resp. $\Gamma \subseteq (0,1]$), whenever $\succeq$ is $Q_3'$-complete (resp. $Q_4'$-complete).

(c)$\Rightarrow$(a). We shall prove only $Q_4''$-completeness (and, hence, $Q_4$-completeness), $Q_3$-completeness can be established in a similar way. Let $\succcurlyeq$ be the preorder on $\mathcal{F}$ induced by $Q_4''$. It is sufficient to prove that for any $\gamma, \sigma \in \Gamma$, $\gamma \geq \sigma \Rightarrow H_\gamma^{(\alpha)} \succcurlyeq H_\sigma^{(\alpha)}$; indeed, if this holds, then $\succcurlyeq$ is complete, and therefore, $\succeq$ is $Q_4''$-complete. Let $\gamma \geq \sigma$ and $x \in Q_4''$ be such that $H_\sigma^{(\alpha)}(x) = x(0) + \sigma(F^{(\alpha)}(x) - x(0)) \geq 0$. Since $x(0) \leq 0$ (as $x \in Q_4''$) and $\sigma > 0$ (as $\Gamma \subseteq (0,1]$), this implies $F^{(\alpha)}(x) - x(0) \geq 0$. Thus, $H_\gamma^{(\alpha)}(x) = x(0) + \gamma(F^{(\alpha)}(x) - x(0)) \geq x(0) + \sigma(F^{(\alpha)}(x) - x(0)) = H_\sigma^{(\alpha)}(x) \geq 0$. ∎

**Proof of Lemma 6.**

Let $\succeq$ be the SPO with a representation $\mathcal{F}$.

(a). Let $\succeq'$ be an $M$-consistent SPO and $\mathcal{F}'$ be a representation of $\succeq'$. Pick $x, y \in Q$. By construction, $M(x) \geq M(y) \Rightarrow x \succeq y$. On the other hand, $\mathcal{F}' \subseteq \mathcal{F}$, so $x \succeq y \Rightarrow x \succeq' y \Rightarrow M(x) \geq M(y)$. Thus, $\succeq$ is the least $M$-consistent SPO.

(b). As $\succeq$ is $M$-consistent and hence Q-complete, $\succcurlyeq$ is complete. Let $\succcurlyeq_D$ be the preorder on $\mathcal{F}$ induced by D. By construction of the set D, $\succcurlyeq \subseteq \succcurlyeq_D$. On the other hand, as $Q \subseteq D$, we have $\succcurlyeq_D \subseteq \succcurlyeq$. Thus, $\succcurlyeq_D = \succcurlyeq$ and $\succcurlyeq_D$ is complete. This proves that $\succcurlyeq_D$ is D-complete.

We only have to show that if $C \supseteq Q$ and $C \setminus D \neq \emptyset$, then $\succeq$ is not C-complete. Let $\succcurlyeq_C$ be the preorder on $\mathcal{F}$ induced by C. Pick $x \in C \setminus D$. By the definition of the set D, it follows that there are $F, G \in \mathcal{F}$ such that $F \succcurlyeq G$, $F(x) < 0$, and $G(x) \geq 0$. The last two inequalities show that it is not true that $F \succcurlyeq_C G$. On the other hand, as $Q \subseteq C$, we have $\succcurlyeq_C \subseteq \succcurlyeq$. As $F \neq G$ and $\succcurlyeq$ is antisymmetric, we get $F \succ G$. This proves that it is not true that $G \succcurlyeq_C F$. Thus, $F$ and $G$ are incomparable with respect to $\succcurlyeq_C$, and therefore, $\succeq$ is not C-complete. ∎

**Proof of Proposition 3.**

(b). Let $\succeq$ be an SPO with a representation $\mathcal{F}$.

Assume that $\succeq$ is $RR^{(A)}$-consistent. Pick $F \in \mathcal{F}$ and denote by $\alpha$ the discount function associated with $F$. Pick $0 < t < \tau$ and set $x_\lambda := -1_0 + (1/\alpha_\lambda(t))1_t$ and $y_\lambda := -1_t + (\alpha_\lambda(t)/\alpha_\lambda(\tau))1_\tau$, $\lambda \in R_+$. Then $x_\lambda, y_\lambda \in Q_2''$ and $RR^{(A)}(x_\lambda) = \lambda = RR^{(A)}(y_\lambda)$. Since $\succeq$ is $RR^{(A)}$-consistent, we must have $I_{\{F\}^\circ}(x_\lambda) = I_{\{F\}^\circ}(y_\lambda)$. The last equality holds for any $\lambda \in R_+$ if and only if there is $\lambda^* \in R_+$ such that $\alpha(t) = \alpha_{\lambda^*}(t)$ and $\alpha(\tau) = \alpha_{\lambda^*}(\tau)$. Therefore, $\mathcal{F} = \{F_\lambda^{(A)}, \lambda \in \Lambda\}$, $\Lambda \subseteq R_+$. $\Lambda$ is dense in $R_+$ (if $R_+ \setminus \Lambda$ contained a proper interval, then it would contradict $RR^{(A)}$-consistency).

To prove the converse assume that $\mathcal{F} = \{F_\lambda^{(A)}, \lambda \in \Lambda\}$, where $\Lambda$ is dense in $R_+$. Clearly, if $x \in Q_2''$, then $\{F \in \mathcal{F} : F(x) \geq 0\} = \{F_\lambda^{(A)}, \lambda \in [0, RR^{(A)}(x)] \cap \Lambda\}$. Therefore, if $x, y \in Q_2''$, then $x \succeq y \Leftrightarrow RR^{(A)}(x) \geq RR^{(A)}(y)$.

(a), (c). These follow from part (b).

(d). The complete preorder $\succcurlyeq$ on $\mathcal{F}^{(A)}$ induced by $Q_2''$ is given by $F_\lambda^{(A)} \succcurlyeq F_{\lambda'}^{(A)} \Leftrightarrow \lambda \leq \lambda'$. Clearly, $\succcurlyeq$ is antisymmetric, so we can use representation (5) for the natural domain. From (5) it follows that $x \in D^{(A)}$ if and only if for any $0 \leq \lambda \leq \lambda'$, $g_x^{(A)}(\lambda') \geq 0 \Rightarrow g_x^{(A)}(\lambda) \geq 0$. That is, the natural domain of $RR^{(A)}$ consists of projects $x \in P$ such that $g_x^{(A)}$ is either nonnegative or negative, or there is $\lambda \in R_+$ such that $g_x^{(A)}$ is nonnegative on $[0, \lambda]$ and negative on $(\lambda, +\infty)$.

(e). Let $\succeq$ be the SPO induced by $\mathcal{F}^{(A)}$. It is straightforward to verify that $\overline{RR}^{(A)}$ is a utility representation for the restriction of $\succeq$ to $D^{(A)}$.

Clearly, $IRR^{(A)}$ is the restriction of $\overline{RR}^{(A)}$ to $Q^{(A)}$, so statements (a)–(e) remain valid with $RR^{(A)}$ replaced by $IRR^{(A)}$. ∎

**Proof of Proposition 4.**

(a)⇒(b). Given a D-family $A$, it is straightforward to verify that the restriction of $IRR^{(A)}$ to $S$ is a rate of return.

(b)⇒(a). Define the function $\phi: R_+ \to R$ by $\phi(z) := M(1, e^z; 0, 1)$ and set $\tilde{M} := \phi^{-1} \circ M$. From conditions 1, 3, and 4 in the definition of a rate of return, it follows that $\tilde{M}$ is well defined and maps $S$ onto $R_+$. The function $\phi^{-1}$ is strictly increasing, so an SPO is $M$-consistent if and only if it is $\tilde{M}$-consistent. In view of Proposition 3 it is sufficient to show that there is a D-family $A$ such that $\tilde{M}(a, b; t, \tau) = RR^{(A)}(1, b/a; t, \tau)$.

Let $J: R_+ \times \{(t, \tau) \in R_+^2 : t < \tau\} \to [1, +\infty)$ be the inverse of $(b; t, \tau) \mapsto \tilde{M}(1, b; t, \tau)$ with respect to the first argument, that is, $\tilde{M}(1, b; t, \tau) = \lambda \Leftrightarrow J(\lambda; t, \tau) = b$. By conditions 1, 3, and 4, $J$ is well defined, and for any $0 \leq t < \tau$, $J(\cdot; t, \tau)$ is strictly increasing and onto $[1, +\infty)$.

Condition 2 implies
$$J(\lambda; t, \tau) J(\lambda; \tau, \delta) = J(\lambda; t, \delta). \tag{10}$$
Extend the domain of $J$ to $R_+^3$ by setting $J(\lambda; t, t) := 1$ and $J(\lambda; \tau, t) := 1/J(\lambda; t, \tau)$ for $0 \leq t < \tau$. Then the Sincov functional equation (10) holds for all $(\lambda, t, \tau, \delta) \in R_+^4$. Its general solution is $J(\lambda; t, \tau) = f(\lambda, t)/f(\lambda, \tau)$ for some function $f: R_+^2 \to R_{++}$ (Aczél, 1966, p. 223). As $J(\lambda; t, \tau) \geq 1$ for all $0 \leq t < \tau$, $f$ is nonincreasing in the second argument. Setting $\alpha_\lambda(t) := f(\lambda, t)/f(\lambda, 0)$, $\lambda \in R_+$, we have $\alpha_\lambda(0) = 1$. Moreover, the function $\lambda \mapsto \alpha_\lambda(t)/\alpha_\lambda(\tau) = J(\lambda; t, \tau)$ is strictly increasing and onto $[1, +\infty)$. Therefore, $\alpha_\lambda$ is a discount function and $A := \langle \alpha_\lambda, \lambda \in R_+ \rangle$ is a D-family. Comparing the definitions of $J$ and $RR^{(A)}$, we conclude that $\tilde{M}(a, b; t, \tau) = RR^{(A)}(1, b/a; t, \tau)$.

(a)⇒(c). Straightforward.

(c)⇒(a). Let $\mathcal{F}$ be a representation of $\succeq$. By Proposition 3 (part (b)), it is sufficient to show that there is a D-family $A$ such that $\mathcal{F} = \{F_\lambda^{(A)}, \lambda \in \Lambda\}$, where $\Lambda$ is a dense subset of $R_+$.

First, we show that each NPV functional in $\mathcal{F}$ is induced by a positive discount function. Assume by way of contradiction that there is $F \in \mathcal{F}$ satisfying $F(1_t) = 0$ for some $t > 0$. By assumption, $-1_t + a1_\tau \succ -1_t + 1_\tau$ for any $\tau > t$ and $a > 1$. This implies that there is a functional $G \in \mathcal{F}$ satisfying $G(1_t) > G(1_\tau)$. Set $x = -1_t + 1_\tau$ and $y = -1_0 + b1_t$, $b \geq 1$. Then $x, y \in Q_2''$, $F(x) = 0$, $G(x) < 0$, whereas $F(y) < 0$, $G(y) \geq 0$ for sufficiently large $b$, so projects $x$ and $y$ are incomparable, which is a contradiction.

Let $\alpha$ and $\beta$ be the discount functions associated with some distinct $F, G \in \mathcal{F}$. Set $\gamma(t) := \alpha(t)/\beta(t)$. As $\succeq$ is $Q_2''$-complete, by Lemma 4, $\gamma$ is monotone. Without loss of generality, we may assume that $\gamma$ is nondecreasing. Let us show that it is actually strictly increasing. Assume by way of contradiction that there are $t_1 < t_2$ such that $\gamma(t_1) = \gamma(t_2)$. Since $\alpha \neq \beta$, there is $\tau$ such that $\gamma(\tau) > 1$. Set $x = -1_0 + (1/\alpha(\tau))1_\tau$ and $y = -1_{t_1} + (\beta(t_1)/\beta(t_2))1_{t_2}$. Then $F(x) = F(y) = G(y) = 0$, $G(x) < 0$, and, therefore, $y \succ x$. Note that $F(-1_{t_1} + b1_{t_2}) < 0$ for any $b < \beta(t_1)/\beta(t_2)$, so the set $U_\succ(x) \cap Q_2''$ does not contain a neighborhood of $y$ in $Q_2''$. As $U_\succ(x) = P \setminus L_\succeq(x)$, this is a contradiction with the lower semicontinuity of the restriction of $\succeq$ to $Q_2''$. This proves that $\alpha(\tau)/\alpha(t) > \beta(\tau)/\beta(t)$ for all $t < \tau$, in particular, $\alpha(\tau) > \beta(\tau)$ for all $\tau > 0$.

The map $F \mapsto -\ln F(1_1)$ defines a bijection between $\mathcal{F}$ and a subset $\Lambda$ of $R_+$. Thus, we can write $\mathcal{F} = \{F_\lambda, \lambda \in \Lambda \subseteq R_+\}$, where $F_\lambda \in \mathcal{F}$ is the NPV functional satisfying $-\ln F_\lambda(1_1) = \lambda$. In what follows the discount function associated with $F_\lambda$ is denoted by $\alpha_\lambda$. By construction, for any $t < \tau$, the function $\lambda \mapsto \alpha_\lambda(\tau)/\alpha_\lambda(t)$ from $\Lambda$ into $(0,1]$ is strictly decreasing. The condition $-1_t + a1_\tau \succ -1_t + b1_\tau$, $t < \tau$, $a > b \geq 1$ implies that for any $t < \tau$, the image of the function $\lambda \mapsto \alpha_\lambda(\tau)/\alpha_\lambda(t)$ is dense in $(0,1]$. In particular, $\Lambda$ is dense in $R_+$ (as the image of the dense subset $\{\alpha_\lambda(1), \lambda \in \Lambda\}$ of $(0,1]$ under the continuous map $z \mapsto -\ln z$).

Let us prove that $\langle \alpha_\lambda, \lambda \in \Lambda \rangle$ can be complemented to a D-family. For each $\lambda \in R_+ \setminus \Lambda$ set $\alpha_\lambda(t) := \sup_{c \in \Lambda: c > \lambda} \alpha_c(t)$. By construction, $\alpha_\lambda$ is a positive discount function. Let us show that $A := \langle \alpha_\lambda, \lambda \in R_+ \rangle$ is a D-family. Given $0 \leq t < \tau$, define the function $\varphi: R_+ \to R_{++}$ by $\varphi(\lambda) := \alpha_\lambda(\tau)/\alpha_\lambda(t)$.

First, we prove that $\varphi$ is strictly decreasing. As $\Lambda$ is dense in $R_+$ and the functions $\lambda \mapsto \alpha_\lambda(t)$ and $\lambda \mapsto \alpha_\lambda(\tau)$ are strictly decreasing on $\Lambda$, we have

$$\varphi(\lambda) = \frac{\alpha_\lambda(\tau)}{\alpha_\lambda(t)} = \frac{\sup_{c \in \Lambda: c > \lambda} \alpha_c(\tau)}{\sup_{c \in \Lambda: c > \lambda} \alpha_c(t)} = \frac{\lim_{c \to \lambda+, c \in \Lambda} \alpha_c(\tau)}{\lim_{c \to \lambda+, c \in \Lambda} \alpha_c(t)} = \lim_{c \to \lambda+, c \in \Lambda} \frac{\alpha_c(\tau)}{\alpha_c(t)} = \lim_{c \to \lambda+, c \in \Lambda} \varphi(c), \quad \lambda \in R_+ \setminus \Lambda.$$

Pick $0 \leq \lambda_1 < \lambda_2$. Since $\Lambda$ is dense in $R_+$, there are $\lambda_1', \lambda_2' \in \Lambda$ such that $\lambda_1 < \lambda_1' < \lambda_2' < \lambda_2$. We have $\varphi(\lambda_1) = \lim_{c \to \lambda_1+, c \in \Lambda} \varphi(c) \geq \varphi(\lambda_1')$, $\varphi(\lambda_2) = \lim_{c \to \lambda_2+, c \in \Lambda} \varphi(c) \leq \varphi(\lambda_2')$, and, therefore, $\varphi(\lambda_1) \geq \varphi(\lambda_1') > \varphi(\lambda_2') \geq \varphi(\lambda_2)$.

To complete the proof we have to show that $\varphi$ is onto $(0,1]$. Assume by way of contradiction that $\varphi(R_+) \neq (0,1]$. Then, since $\varphi$ is monotone, $(0,1] \setminus \varphi(R_+)$ contains an interval of positive length, which is a contradiction with density of $\varphi(\Lambda)$ in $(0,1]$. ∎

**Proof of Proposition 5.**
(a) $\Rightarrow$ (b). Trivial.
(b) $\Rightarrow$ (a). Let $\mathcal{F}$ be a representation of $\succeq$. In view of Proposition 3 (part (b)), we have to show that $\mathcal{F} = \{F_\lambda^{(E)}, \lambda \in \Lambda\}$, where $\Lambda$ is a dense subset of $R_+$. Pick $F \in \mathcal{F}$ and denote by $\alpha$ the discount function associated with $F$. Note that if $x \in Q_2''$, then so does $x^{(+\tau)}$.

First, we prove that $\alpha$ is positive. Assume by way of contradiction that $\alpha(\tau) = 0$ for some $\tau > 0$. Consider project $x = -1_0 + a1_\tau \in Q_2''$. Then $F(x) < 0$, whereas $F(x^{(+\tau)}) = 0$, a contradiction to stationarity.

Pick $t > 0$ and consider project $y = -1_0 + c1_t \in Q_2''$ for some $c \geq 1$. If $c = 1/\alpha(t)$, then $F(y) = -1 + c\alpha(t) = 0$, and, by stationarity, we must have
$$-\alpha(\tau) + (1/\alpha(t))\alpha(t+\tau) = -\alpha(\tau) + c\alpha(t+\tau) = F(-1_\tau + c1_{t+\tau}) = F(y^{(+\tau)}) \geq 0 \tag{11}$$

for any $\tau > 0$. We consider two cases.

*Case 1*: $\alpha(t) < 1$. Setting $c \in [1, 1/\alpha(t))$, a similar argument used to obtain (11) yields $-\alpha(\tau) + c\alpha(t + \tau) < 0$. Tending $c \to 1/\alpha(t)-$ in the last inequality and combining the result with (11), we obtain the Cauchy functional equation $\alpha(t + \tau) = \alpha(t)\alpha(\tau)$, $(t, \tau) \in \mathrm{R}_{++}^2$. Its general positive nonincreasing solution is given by $\alpha(t) = e^{-\lambda t}$, $\lambda \in \mathrm{R}_+$ (Aczél, 1966, p. 38).

*Case 2*: $\alpha(t) = 1$. Using (11) with $\tau = t$, we get $\alpha(2t) = 1$. Iterating this equality, we conclude that $\alpha = 1_0$.

Thus, there exists $\Lambda \subseteq \mathrm{R}_+$ such that $\mathcal{F} = \{F_\lambda^{(\mathrm{E})}, \lambda \in \Lambda\}$. Pick $a > b \geq 1$. By assumption, there exists $t > 0$ such that $-1_0 + a1_t \succ -1_0 + b1_t$. Therefore, there is $\lambda \in \Lambda$ such that $\lambda \in (t^{-1} \ln b, t^{-1} \ln a]$. This proves that $\Lambda$ is dense in $\mathrm{R}_+$. ∎

**Proof of Lemma 7.**

For any $\mathcal{S} \in \mathrm{P}^*$ and $z \in \mathrm{P}$, set $\mathcal{S}(z) := \{z\}^\circ \cap \mathcal{S}$.

We shall prove a slightly stronger result: if $x \in \mathcal{R}(\mathcal{F})$, then $\mathrm{U}_\succeq(x) = \mathrm{U}_{\succeq'}(x)$. During the proof cl and int are the topological closure and interior operators in $\mathcal{F}$. Clearly, $\mathrm{U}_\succeq(x) \subseteq \mathrm{U}_{\succeq'}(x)$. To prove the converse, we have to show that for any $y \in \mathrm{P}$, $\mathcal{F}'(x) \subseteq \mathcal{F}'(y) \Rightarrow \mathcal{F}(x) \subseteq \mathcal{F}(y)$. Note that $\mathcal{F}'(x) \subseteq \mathcal{F}'(y)$ implies $\mathrm{cl}(\mathcal{F}(x) \cap \mathcal{F}') = \mathrm{cl}(\mathcal{F}'(x)) \subseteq \mathrm{cl}(\mathcal{F}'(y)) = \mathrm{cl}(\mathcal{F}(y) \cap \mathcal{F}') \subseteq \mathcal{F}(y)$, where the last inclusion follows from the fact that $\mathcal{F}(y)$ is closed in $\mathcal{F}$ (as the intersection of the closed half-space $\{y\}^\circ$ and $\mathcal{F}$). Therefore, it is sufficient to prove that $\mathcal{F}(x) \subseteq \mathrm{cl}(\mathcal{F}(x) \cap \mathcal{F}')$, which readily follows from regularity of $x$. Indeed, let $O$ be an open (in $\mathcal{F}$) neighborhood of a point from $\mathrm{int}\,\mathcal{F}(x)$. Since $O \cap \mathrm{int}\,\mathcal{F}(x)$ is a nonempty open set and $\mathcal{F}'$ is dense in $\mathcal{F}$, $O \cap \mathrm{int}\,\mathcal{F}(x)$ intersects $\mathcal{F}'$ and therefore also intersects $\mathcal{F}(x) \cap \mathcal{F}'$ (as $\mathrm{int}\,\mathcal{F}(x) \subseteq \mathcal{F}(x)$). Thus, every neighborhood of a point from $\mathrm{int}\,\mathcal{F}(x)$ intersects $\mathcal{F}(x) \cap \mathcal{F}'$, so $\mathrm{int}\,\mathcal{F}(x) \subseteq \mathrm{cl}(\mathcal{F}(x) \cap \mathcal{F}')$. Taking the closure of both sides, we get $\mathcal{F}(x) = \mathrm{cl}(\mathrm{int}\,\mathcal{F}(x)) \subseteq \mathrm{cl}(\mathcal{F}(x) \cap \mathcal{F}')$ due to regularity of $x$. ∎

**Lemma 10.**

*Given a D-family* $\mathrm{A}$, *the map* $h(\lambda) := F_\lambda^{(\mathrm{A})}$ *is a homeomorphism between* $\mathrm{R}_+$ *and* $\mathcal{F}^{(\mathrm{A})}$ *endowed with the subspace topology.*

**Proof.**

Clearly, $h$ is a bijection. Pick $\lambda^* \in \mathrm{R}_+$.

Consider a convergent sequence $\lambda_n \to \lambda^*$. For any $x \in \mathrm{P}$ the function $\lambda \mapsto F_\lambda^{(\mathrm{A})}(x)$ is continuous, $F_{\lambda_n}^{(\mathrm{A})} \to F_{\lambda^*}^{(\mathrm{A})}$ pointwise, so $h$ is continuous at $\lambda^*$.

In order to prove that $h^{-1}$ is continuous at $F_{\lambda^*}^{(\mathrm{A})}$, pick $t > 0$ and recall that by the definition of a D-family, the function $\lambda \mapsto \alpha_\lambda(t)$ is a homeomorphism of $\mathrm{R}_+$ onto $(0, 1]$. In particular, as its inverse is continuous, for any $\varepsilon > 0$, there is $\delta > 0$ such that $|\alpha_\lambda(t) - \alpha_{\lambda^*}(t)| < \delta \Rightarrow |\lambda - \lambda^*| < \varepsilon$. As $\{F_\lambda^{(A)} : |F_\lambda^{(A)}(1_t) - F_{\lambda^*}^{(A)}(1_t)| < \delta\} = \{F_\lambda^{(A)} : |\alpha_\lambda(t) - \alpha_{\lambda^*}(t)| < \delta\}$ is an open neighborhood of $F_{\lambda^*}^{(A)}$ in $\mathcal{F}^{(A)}$, we are done. ∎

**Proof of Proposition 6.**

(b). Let $\succeq$ be an SPO with a representation $\mathcal{F}$.

Assume that $\succeq$ is $RDPP^{(\alpha)}$-consistent. Pick $F \in \mathcal{F}$ and denote by $\beta$ be the discount function associated with $F$. Pick $\tau > 0$.

*Claim 1*: if $\alpha(\tau) = 0$, then $\beta(\tau) = 0$. To see this note that since $\alpha \neq \chi$, there is $t \in (0, \tau)$ such that $\alpha(t) > 0$. Set $x = -1_0 + (1/\alpha(t))1_t$. Then $DPP^{(\alpha)}(x) = DPP^{(\alpha)}(x + c1_\tau) = t$ for each $c \in \mathbb{R}$. Thus, $x \sim x + c1_\tau$ and we must have $I_{\{F\}^\circ}(x) = I_{\{F\}^\circ}(x + c1_\tau)$. The last equality holds for all $c \in \mathbb{R}$ if and only if $\beta(\tau) = 0$.

*Claim 2*: if $\beta(\tau) > 0$, then $\beta(\tau) \geq \alpha(\tau)$. Indeed, assume by way of contradiction that $\alpha(\tau) > \beta(\tau)$ and consider projects $x = -1_0 + (1/\alpha(\tau))1_\tau$ and $y = -1_0 + (1/\beta(\tau))1_\tau$. Then $DPP^{(\alpha)}(x) = DPP^{(\alpha)}(y) = \tau$, whereas $F(x) < 0$ and $F(y) = 0$, so it is not true that $x \sim y$; a contradiction with $RDPP^{(\alpha)}$-consistency.

*Claim 3*: if $\beta(\tau) > 0$, then there is $\lambda \geq 1$ such that $\beta(t) = \lambda \alpha(t)$ for all $t \in (0, \tau]$. Indeed, pick $t \in (0, \tau)$. By Claims 1 and 2, $\beta(\tau) \geq \alpha(\tau) > 0$ and $\beta(t) \geq \alpha(t) > 0$. Consider projects $x = -1_0 + (1/\alpha(\tau))1_\tau$, $y = -1_0 + (1/\alpha(t) - \varepsilon)1_t + \varepsilon \alpha(t)/\alpha(\tau) 1_\tau$. Then $DPP^{(\alpha)}(x) = DPP^{(\alpha)}(y) = \tau$ for any $\varepsilon > 0$. Thus, $x \sim y$ and we must have $I_{\{F\}^\circ}(x) = I_{\{F\}^\circ}(y)$. In view of Claim 2, $I_{\{F\}^\circ}(x) = 1$. The equality $I_{\{F\}^\circ}(y) = 1$ holds for all $\varepsilon > 0$ if and only if $\beta(t)/\alpha(t) \leq \beta(\tau)/\alpha(\tau)$. By considering projects $x' = -1_0 + (1/\alpha(t))1_t$ and $y' = -1_0 + (1/\alpha(t) + \varepsilon)1_t - \varepsilon \alpha(t)/\alpha(\tau) 1_\tau$ with $\varepsilon > 0$, in the same manner we arrive at $\beta(t)/\alpha(t) \geq \beta(\tau)/\alpha(\tau)$.

*Claim 4*: if $\beta(\tau) = \lambda \alpha(\tau) > 0$, $\lambda > 1$, then $\beta(t) = \lambda \alpha(t)$ for all $t \in \mathbb{R}_{++}$. Pick $t \in \mathbb{R}_{++}$. If $t \in (0, \tau]$, the statement follows from Claim 3. Now assume that $t > \tau$. If $\alpha(t) = 0$, the statement follows from Claim 1. If $\alpha(t) > 0$, then, by Claim 3, either $\beta(t) = \lambda \alpha(t)$ or $\beta(t) = 0$. Assume that $\beta(t) = 0$ and consider projects $x = -1_0 + (1/\beta(\tau))1_\tau + c1_t$ and $y = -1_0 + c1_t$. Then $DPP^{(\alpha)}(x) = DPP^{(\alpha)}(y) = t$ for sufficiently large $c$, whereas $F(x) = 0$ and $F(y) < 0$ for any $c$; a contradiction with $RDPP^{(\alpha)}$-consistency. This proves that $\beta(t) = \lambda \alpha(t)$.

*Claim 5*: if $\beta(\tau) = \alpha(\tau) > 0$, then either $\beta = \alpha$ or there is $c \in [\tau, +\infty)$ such that $\beta(t) = \begin{cases} \alpha(t) & \text{if } t \leq c \\ 0 & \text{if } t > c \end{cases}$. From Claims 2 and 3 it follows that either $\beta = \alpha$, or $\beta(t) = \begin{cases} \alpha(t) & \text{if } t \leq c \\ 0 & \text{if } t > c \end{cases}$, $c \in [\tau, +\infty)$, or $\beta(t) = \begin{cases} \alpha(t) & \text{if } t < c \\ 0 & \text{if } t \geq c \end{cases}$, $c \in (\tau, +\infty)$. Unless $\alpha(t) = 0$, the latter opportunity contradicts $RDPP^{(\alpha)}$-consistency: consider projects $x$ and $y$ such that $DPP^{(\alpha)}(x) = DPP^{(\alpha)}(y) = c$ and the function $t \mapsto G_t^{(\alpha)}(x)$ (resp. $t \mapsto G_t^{(\alpha)}(y)$) is continuous at $c$ (resp. $\lim_{t \to c-} G_t^{(\alpha)}(y) < 0$), then $F(x) = 0$, but $F(y) < 0$.

*Claim 6*: $\beta \neq \chi$. Since $\alpha \neq \chi$, there is $\tau'$ such that $\alpha(\tau') > 0$. Assume by way of contradiction that $\beta = \chi$ and consider projects $x(t) = (1_{\tau'}(t) - 1)t + c1_{\tau'}(t)$ and $y = -1_0 + c1_{\tau'}$. Then $DPP^{(\alpha)}(x) = DPP^{(\alpha)}(y) = \tau'$ for sufficiently large $c$, whereas $F(x) = 0$ and $F(y) < 0$ for any $c$, which is a contradiction.

Combining Claims 1–6 we get that $\mathcal{F} = \{G_\tau^{(\alpha)}, \tau \in T\} \cup \{H_\gamma^{(\alpha)}, \gamma \in \Gamma\}$ for some $T \subseteq R_{++}$ and $\Gamma \subseteq [1, 1/\alpha(0+)]$. Without loss of generality, we may assume that $T \subseteq \text{int}(\text{supp}\{\alpha\})$. Clearly, T must be dense in $\text{int}(\text{supp}\{\alpha\})$ in order to $\succeq$ be $RDPP^{(\alpha)}$-consistent.

To prove the converse, assume that $\mathcal{F} = \{G_\tau^{(\alpha)}, \tau \in T\} \cup \{H_\gamma^{(\alpha)}, \gamma \in \Gamma\}$, where T is a dense subset of $\text{int}(\text{supp}\{\alpha\})$ and $\Gamma \subseteq [1, 1/\alpha(0+)]$. Clearly, if $x \in Q^{(\alpha)}$, then $\{F \in \mathcal{F} : F(x) \geq 0\} = \{G_\tau^{(\alpha)}, \tau \in [DPP^{(\alpha)}(x), +\infty) \cap T\} \cup \{H_\gamma^{(\alpha)}, \gamma \in \Gamma\}$. Therefore, if $x, y \in Q^{(\alpha)}$, then $x \succeq y \Leftrightarrow DPP^{(\alpha)}(x) \leq DPP^{(\alpha)}(y) \Leftrightarrow RDPP^{(\alpha)}(x) \geq RDPP^{(\alpha)}(y)$.

(a), (c). These follow from part (b).

(d). Let $\succeq$ be the least $RDPP^{(\alpha)}$-consistent SPO. By part (c) it is induced by $\mathcal{F} = \{G_\tau^{(\alpha)}, \tau \in \text{int}(\text{supp}\{\alpha\})\} \cup \{H_\gamma^{(\alpha)}, \gamma \in [1, 1/\alpha(0+)]\}$. Note that unless $\alpha(0+) = 1$, the preorder on $\mathcal{F}$ induced by $Q^{(\alpha)}$ is not antisymmetric, so we cannot use Lemma 6 to derive the natural domain.

Pick $x \in P$. We consider three cases.

*Case 1*: $x \in Q_{-}^{(\alpha)}$. In this case, the restriction of $\succeq$ to $Q^{(\alpha)} \cup \{x\}$ is complete.

*Case 2*: there are $t, \tau \in R_{++}$ such that $G_t^{(\alpha)} \geq 0$ and $G_\tau^{(\alpha)} < 0$. Then one can show that in order to the restriction of $\succeq$ to $Q^{(\alpha)} \cup \{x\}$ be complete we must have $x \in Q^{(\alpha)}$.

*Case 3*: $G_\tau^{(\alpha)}(x) \geq 0$ for all $\tau \in R_{++}$. Then in order to the restriction of $\succeq$ to $Q^{(\alpha)} \cup \{x\}$ be complete we must have $H_\gamma^{(\alpha)}(x) \geq 0$ for all $\gamma \in [1, 1/\alpha(0+)]$, or, equivalently, $x \in Q_{+}^{(\alpha)}$.

It is straightforward to verify that the restriction of $\succeq$ to $Q_{-}^{(\alpha)} \cup Q^{(\alpha)} \cup Q_{+}^{(\alpha)}$ is complete.

(e). Let $\succeq$ be the SPO induced by $\{G_\tau^{(\alpha)}, \tau \in \text{int}(\text{supp}\{\alpha\})\} \cup \{H_\gamma^{(\alpha)}, \gamma \in [1, 1/\alpha(0+)]\}$. It is routine to verify that $\overline{RDPP}^{(\alpha)}$ is a utility representation for the restriction of $\succeq$ to $D^{(\alpha)}$. ∎

**Proof of Proposition 7.**

(a)$\Rightarrow$(b). First, we show that if $x = -1_0 + a1_\tau \in Q_1'$, $y = -1_0 + b1_t \in Q_1'$, $0 < \tau < t$, and $x \succ -1_0$, then $x \succ y$ (during the proof, we refer to this implication as (*)). Indeed, assume by way of contradiction that $x \succ y$ does not hold. Then, by $Q_1'$-completeness, $y \succeq x$. Using stability under truncation, we get $-1_0 = y_{\leq \tau} \succeq x_{\leq \tau} = x$, which is a contradiction.

Set $\alpha(t) := \sup_{F \in \mathcal{F}} F(1_t)$. Clearly, $\alpha \in \mathcal{A}$. If $\alpha = \chi$, then $\mathcal{F} = \{F^{(\chi)}\}$ and (b) holds trivially. So in what follows we assume that $\alpha \neq \chi$. Let $\beta$ be the discount function associated with some $G \in \mathcal{F}$. Pick $\tau > 0$ and assume that $\alpha(\tau) > \beta(\tau) > 0$. Setting $x = -1_0 + a1_\tau$ with $a \in (1/\alpha(\tau), 1/\beta(\tau))$, we have $x \succ -1_0$ and $G(x) < 0$. Then, for any $t > \tau$ and $b \geq 1$, (*) implies $G(-1_0 + b1_t) < 0$. This proves that $\beta(t)=0$ for all $t>\tau$. Thus, $\mathcal{F} \subseteq \{F^{(\chi)}, F^{(\alpha)}\} \cup \{G_{t,\lambda}^{(\alpha)}, (t,\lambda) \in (\text{supp}\{\alpha\}\setminus\{0\})\times[0,1]\}$. By (*), the set $T := \{t \in R_{++}: \text{there is } \lambda \in [0,1] \text{ such that } G_{t,\lambda}^{(\alpha)} \in \mathcal{F}\}$ is dense in $\text{supp}\{\alpha\}$.

We claim that $G_\tau^{(\alpha)} \in \mathcal{F}$ for any $\tau \in \{0\} \cup \text{int}(\text{supp}\{\alpha\})$. Indeed, pick $\tau \in \{0\} \cup \text{int}(\text{supp}\{\alpha\})$ and assume by way of contradiction that $G_\tau^{(\alpha)} \notin \mathcal{F}$. One can construct a project $x$ such that $G_\tau^{(\alpha)}(x) = 0$, $F(x) < 0$ for any $F \in \{F^{(\chi)}, F^{(\alpha)}\} \cup \{G_{t,\lambda}^{(\alpha)}, (t,\lambda) \in R_{++} \times [0,1]\} \setminus \{G_\tau^{(\alpha)}\}$, and therefore, $x \sim -1_0$. By stability under truncation, $x_{\leq \tau} \sim (-1_0)_{\leq \tau} = -1_0$. Since T is dense in $\text{supp}\{\alpha\}$, there are $t \in T$ and $\lambda \in [0,1]$ such that $t > \tau$ and $G_{t,\lambda}^{(\alpha)} \in \mathcal{F}$. We have $G_{t,\lambda}^{(\alpha)}(x_{\leq \tau}) = G_\tau^{(\alpha)}(x) = 0$, which is a contradiction to $x_{\leq \tau} \sim -1_0$. This proves that $G_\tau^{(\alpha)} \in \mathcal{F}$.

(b)$\Rightarrow$(a). Clearly, $\mathcal{F}$ is completely ordered by $\succcurlyeq_1$, so $\succeq$ is $Q_1'$-complete (Lemma 3). To establish stability under truncation, it is sufficient to prove that for any $F \in \mathcal{F}$ and $\tau \in R_+$, the NPV functional $x \mapsto F(x_{\leq \tau})$ also belongs to $\mathcal{F}$. Pick $x \in P$ and $\tau \in R_+$. We have $F^{(\chi)}(x_{\leq \tau}) = F^{(\chi)}(x)$; $F^{(\alpha)}(x_{\leq \tau}) = G_\tau^{(\alpha)}(x)$ if $\tau \in \{0\} \cup \text{int}(\text{supp}\{\alpha\})$; $F^{(\alpha)}(x_{\leq \tau}) = F^{(\alpha)}(x)$ if $\tau \notin \{0\} \cup \text{int}(\text{supp}\{\alpha\})$; $G_{t,\lambda}^{(\alpha)}(x_{\leq \tau}) = G_{t,\lambda}^{(\alpha)}(x)$ if $0 < t \leq \tau$; $G_{t,\lambda}^{(\alpha)}(x_{\leq \tau}) = G_\tau^{(\alpha)}(x)$ if $t > \tau$ (note that the conditions $G_{t,\lambda}^{(\alpha)} \in \mathcal{F}$ and $t > \tau$ imply that $\tau \in \{0\} \cup \text{int}(\text{supp}\{\alpha\})$). This completes the proof. ∎

**Proof of Proposition 8.**

(b). Let $\succeq$ be an SPO with a representation $\mathcal{F}$.

Assume that $\succeq$ is $RI_G^F$-consistent. Pick $H \in \mathcal{F}$ and $x \in Q_G^F$. Consider $y \in P$ such that $F(y) = G(y) = 0$. For any $\lambda \in R$, we have $x + \lambda y \in Q_G^F$ and $RI_G^F(x) = RI_G^F(x + \lambda y)$. By $RI_G^F$-consistency, we must have $I_{\{H\}^\circ}(x) = I_{\{H\}^\circ}(x + \lambda y)$. The last equality holds for all $\lambda \in R$ if and only if $H(y) = 0$. This implies that $H = aG + bF$ for some scalars $a$ and $b$ (Aliprantis and Border, 2006, Lemma 5.91, p. 212). As $H \in \mathcal{F}$, we have $a + b = 1$. Therefore, $\mathcal{F} = \{wF + (1-w)G, w \in W\}$, $W \subseteq \tilde{W}$. $W \cap (0,1)$ is dense in $(0,1)$ (indeed, if $(0,1) \setminus W$ contained a proper interval, then it would contradict $RI_G^F$-consistency). Density of $W \cap (0,1)$ in $(0,1)$ implies that $\mathcal{F}$ and $\{wF + (1-w)G, w \in W \cup (0,1)\}$ represent the same SPO.

To prove the converse, assume that $\mathcal{F} = \{wF + (1-w)G, w \in W\}$, $(0,1) \subseteq W \subseteq \tilde{W}$. If $x \in Q_G^F$, then $\{x\}^\circ \cap \mathcal{F} = \{wF + (1-w)G, w \in [1/RI_G^F(x), +\infty) \cap W\}$. Therefore, provided that $x, y \in Q_G^F$, $x \succeq y \Leftrightarrow RI_G^F(x) \geq RI_G^F(y)$.

(a), (c). These follow from part (b).

(d). Let $\succeq$ be the least $RI_G^F$-consistent SPO. By part (c) it is induced by $\{w\tilde{F} + (1-w)\tilde{G}, w \in [0,1]\}$. It is straightforward to verify that the restriction of $\succeq$ to $D_G^F$ is complete. On the other hand, if $x \notin D_G^F$, i.e., $\tilde{F}(x) < 0$ and $\tilde{G}(x) \geq 0$, then $x$ and $y \in Q_G^F$ are incomparable (recall that $\tilde{F}(y) \geq 0$ and $\tilde{G}(y) < 0$ for all $y \in Q_G^F$).

(e). Let $\succeq$ be the SPO induced by $\{w\tilde{F} + (1-w)\tilde{G}, w \in [0,1]\}$. It is straightforward to verify that $\overline{RI}_G^F$ is a utility representation for the restriction of $\succeq$ to $D_G^F$. ∎

**Lemma 11.**

For the function $\pi: Q_1' \to [1, +\infty)$ defined by $\pi(-1_0 + b1_\tau) := b$, the following statements hold.

(a) $\pi$ is a profitability metric.

(b) An SPO is $\pi$-consistent if and only if it is induced by $\{H_\gamma^{(1_0)}, \gamma \in \Gamma\}$, where $(0,1) \subseteq \Gamma \subseteq [0,1]$.

(c) The least $\pi$-consistent SPO is induced by $\{H_\gamma^{(1_0)}, \gamma \in [0,1]\}$.

(d) The natural domain of $\pi$ is $D = P \setminus \{x \in P: x(0) \geq 0, x(+\infty) < 0\}$.

(e) The natural extension of $\pi$ is the complete preorder on $D$ with a utility representation $\bar{\pi}: D \to \bar{R}_+$ given by (6).

**Proof.**

It is sufficient to prove part (b). The rest then follows from Proposition 8 with $F = F^{(1_0)}$ and $G = F^{(\chi)}$.

Let $\succeq$ be a $\pi$-consistent SPO and $\mathcal{F} \subseteq \mathcal{NPV}$ represent $\succeq$. Pick $F \in \mathcal{F}$ and denote by $\alpha$ the discount function associated with $F$. Choose $t, \tau \in R_{++}$. As $\pi(-1_0 + b1_\tau) = \pi(-1_0 + b1_t)$, $b \geq 1$, by $\pi$-consistency, we have $I_{R_+}(-1 + b\alpha(\tau)) = I_{\{F\}^\circ}(-1_0 + b1_\tau) = I_{\{F\}^\circ}(-1_0 + b1_t) = I_{R_+}(-1 + b\alpha(t))$. This equality holds for all $b \geq 1$ if and only if $\alpha(\tau) = \alpha(t)$. Therefore, $\mathcal{F} = \{H_\gamma^{(1_0)}, \gamma \in \Gamma\}$, $\Gamma \subseteq [0,1]$. $\Gamma \cap (0,1)$ is dense in $(0,1)$ (indeed, if $(0,1) \setminus \Gamma$ contained a proper interval, then it would contradict $\pi$-consistency). Density of $\Gamma \cap (0,1)$ in $(0,1)$ implies that $\mathcal{F}$ and $\{H_\gamma^{(1_0)}, \gamma \in \Gamma \cup (0,1)\}$ represent the same SPO.

It is straightforward to verify that the SPO induced by $\{H_\gamma^{(1_0)}, \gamma \in \Gamma\}$, $(0,1) \subseteq \Gamma \subseteq [0,1]$ is $\pi$-consistent. ∎